\documentclass[reprint,amsmath,amssymb,aps,prl,superscriptaddress]{revtex4-1}

\usepackage{bm}
\newcommand{\ii}{\mathrm{i}}

\newcommand{\vq}{{\bf{q}}}
\newcommand{\vk}{{\bf{k}}}

\usepackage{graphicx}
\usepackage{comment}

\usepackage[
	colorlinks,
	breaklinks,]{hyperref}

\usepackage{color}
\usepackage{soul}
\usepackage{ulem}

\begin{document}

\title{
Effects of quantum fluctuations 
on the low-energy collective modes \\
of two-dimensional superfluid Fermi gases from 
the BCS to the Bose Limit
}
%
%

\author{Senne Van Loon}
\email{Senne.Van\_Loon@colostate.edu}
\affiliation{School of Physics, Georgia Institute of Technology, Atlanta, Georgia 30332, USA}
\affiliation{TQC, Universiteit Antwerpen, Universiteitsplein 1, B-2610
                Antwerpen, Belgi\"e}
\author{C. A. R. S\'a de Melo}
\affiliation{School of Physics, Georgia Institute of Technology, Atlanta, Georgia 30332, USA}

\date{\today}

\begin{abstract}
We investigate the effects of quantum fluctuations on the low-energy collective
modes of two-dimensional (2D) $s$-wave Fermi superfluids from the BCS to the
Bose limit. We compare our results to recent Bragg scattering experiments in 2D
box potentials, with very good agreement. We show that quantum fluctuations in
the phase and modulus of the pairing order parameter are absolutely necessary to
give physically acceptable chemical potential and dispersion relation of the
low-energy collective mode throughout the BCS to Bose evolution. Furthermore, we
demonstrate that the dispersion of the collective modes change from concave to
convex as interactions are tuned from the BCS to the Bose regime, and never
crosses the two-particle continuum, because arbitrarily small attractive
interactions produce bound states in 2D.
\end{abstract}

\maketitle

The study of collective modes is a fundamental component of many-particle
physics, because for every spontaneously broken continuous symmetry there are
low-energy modes that emerge as expected from Goldstone's
theorem~\cite{Goldstone1961}, and additional higher-energy excitations such as
the Higgs mode~\cite{Higgs1966, Englert1966}. Collective modes are essential in
understanding a variety of systems ranging from condensed matter (quantum
magnets, superconductors)~\cite{Tennant2017, Hisrchfeld2015}, high energy
physics (standard nuclear matter, quantum chromodynamics)~\cite{Matera2003,
Baym2007}, and astrophysics (neutron stars, black holes)~\cite{Pethick2013,
Turton2014} to atomic (Bose and Fermi superfluids) physics~\cite{Mathey2021,
Hazzard2020}. Unfortunately, in condensed matter it is not easy to tune
parameters such as interactions, density, and dimensionality over a wide range,
in high energy physics it is very difficult, and in astrophysics it is
impossible. However, in atomic physics this is relatively easy via well
established techniques~\cite{Chin2010, Hulet2020}. This makes it possible to
investigate collective modes in superfluids, particularly important because they
reveal the effects of quantum fluctuations above the superfluid ground state.

Superfluids in 2D are inherently different from their 3D counterparts, due to
the importance of fluctuations~\cite{Mermin1966,Hohenberg1967} leading to a
Berezinskii-Kosterlitz-Thouless (BKT) transition~\cite{Berezinskii1972,
Kosterlitz1972}. In the context of ultracold atoms, the desire to study 2D Fermi
superfluids is driven not only by connections to high-temperature
superconductors~\cite{Keimer2015, Ge2015, Cao2018, Yu2019}, but also by the high
degree of experimental control that allows the measurement of the equation of
state~\cite{Turlapov2014, Vale2016, Enss2016}, the observation of the BKT
transition~\cite{Hadzibabic2006,Jochim2015}, and the examination of collective
modes~\cite{Dalibard2018, Henning2020, Christodoulou2021, Lompe2021}.

Ultracold fermions with tunable interactions in nearly 2D configurations were
studied using harmonic traps and optical
lattices~\cite{Turlapov2010,Feld2011,Vale2011,Thomas2015}. With the very recent
advent of box potentials, it is now possible to study experimentally homogeneous
2D fermions~\cite{Henning2018, Henning2020, Lompe2021}. Inspired by recent
measurements of collective excitations via
Bragg-spectroscopy~\cite{Hoinka2017,Biss2021,Lompe2021}, we investigate the
low-energy collective modes of 2D $s$-wave Fermi superfluids in box potentials, and find
very good agreement with experiments. We establish that mean field (saddle
point) results in 2D~\cite{Randeria1990} produce incorrect values of the
chemical potential and lead to the erroneous conclusion that the sound velocity
is a constant throughout the BCS to Bose evolution~\cite{Marini1998,
Salasnich2013, Lumbeeck2020}. In sharp contrast, we show that the inclusion of
quantum fluctuations~\cite{HuLiu2015} is crucial to produce physically
acceptable results for the dispersion of collective modes and leads to a varying
speed of sound in the crossover from BCS to Bose regimes at low
temperatures~\cite{Salasnich2016}. Furthermore, we demonstrate that phase and
modulus fluctuations of the pairing order parameter become increasingly more
coupled with growing interaction strength. Importantly, we clarify the
longstanding confusion about the difference between the resulting sound mode
arising from the broken U(1) symmetry and Landau's phenomenological first sound. 

Based on weakly coupled $s$-wave Fermi superfluids and a linear dispersion of the
collective mode, it has been long thought~\cite{Popov1991} that the low-energy
collective modes in neutral Fermi superfluids are strongly damped (due to Landau
damping) when the energy of the collective mode is sufficiently large to reach
the pairbreaking energy threshold. In 3D, this view is still valid, even when
taking into account the changing concavity of the
dispersion~\cite{Combescot2006,Kurkjian2016}. However, we show that the
situation is fundamentally different in 2D, where the inclusion of the
ubiquitous bound states and of higher-order momentum corrections to the
collective mode dispersion show that the collective mode energy never reaches
the two-particle continuum, and thus there is no damping of the collective mode
at the Gaussian level for $s$-wave superfluids.

{\it Hamiltonian:}
To analyze the low-energy collective modes of 2D $s$-wave Fermi superfluids in box
potentials, we start from the Hamiltonian density
\begin{equation}
\label{eqn:hamiltonian}
{\cal H} = 
\psi^\dagger_s ({\bf r})
\frac{(-i \hbar \nabla)^2}{2m}
\psi_s ({\bf r})
-
g 
\psi^\dagger_\uparrow ({\bf r}) 
\psi^\dagger_\downarrow ({\bf r})
\psi_\downarrow ({\bf r})
\psi_\uparrow ({\bf r}),
\end{equation}
where $\psi_s ({\bf r})$ is a fermion field operator with spin $s$ at position
${\bf r}$. The first term is the kinetic energy and the second represents local
attractive interactions. The associated action is
$
{\cal S} (\psi^\dagger, \psi)
= 
\int d^3r 
\{
\psi^\dagger (r) 
\left[ \hbar \partial_{\tau} - \mu \right]
\psi(r)
+ {\cal H} (r)
\},
$
where $r = ({\bf r}, \tau),\, \int d^3r = \int_0^{\beta} d\tau \int d^2{\bf
r},~\, \beta = \hbar/k_B T$, and $\mu$ is the chemical potential. The grand
canonical partition function of the system is ${\cal Z} = \int D\psi^\dagger
D\psi e^{-{\cal S}/\hbar}$. We introduce the Hubbard-Stratonovich complex pair
field $\Phi (r)$ to decouple the contact interactions and integrate out the
fermionic fields to obtain an effective action $S_{\rm eff} (\Phi^\dagger,
\Phi)$. We write $\Phi (r) = |\Phi (r)| e^{i \theta (r)}$ in terms of its
modulus $\vert \Phi (r) \vert =  \vert \Delta \vert \left[ 1 + \lambda (r)
\right]$ and phase $\theta (r)$, and expand $S_{\rm eff} (\Phi^\dagger, \Phi)$
up to quadratic order in both the phase $\theta (r)$ and modulus fluctuations
$\lambda (r)$ around the saddlepoint $\vert \Phi_{\rm sp} (r) \vert = \vert
\Delta \vert$. The resulting Gaussian action is
\begin{equation}
    S_{\rm eff} = S_{\rm sp} + \beta \vert \Delta \vert^2 
    \sum_{q} 
    \begin{pmatrix}
        \ii \theta_{-q} & \lambda_{-q}
    \end{pmatrix}
    M(q)
    \begin{pmatrix}
        -\ii \theta_{q} \\ \lambda_{q}
    \end{pmatrix},
    \label{eqn:action-gaussian}
\end{equation}
where $q = ({\bf q}, \ii\nu)$ and $\nu = 2\pi n/\beta$ are bosonic Matsubara
frequencies, $S_{\rm sp} = \beta \sum_\vk (\xi_\vk-E_\vk) + \beta
L^2\vert\Delta\vert^2/g$ is the saddlepoint action, and $M(q)$ is the symmetric
Gaussian fluctuation matrix~\cite{SaDeMelo1997}. Using the analytic continuation
$\ii\nu \to \omega + \ii\delta$, the matrix elements of $M (q)$, at zero
temperature, are
\begin{align}
\frac{M_{\!_{\pm \pm}}}{L^2} 
= & \int\!\frac{d^2{\bf k}}{4\pi^2}\!
\left[ 
\frac{E_{+}+ E_{-}}
{2E_{+}E_{-}}
\frac{E_{+}E_{-}
+\xi_{+}\xi_{-} \pm \vert\Delta\vert^2}
{\hbar^2\omega^2 - (E_{+} + E_{-})^2} 
\! +\! \frac{1}{2 E_{\bf k}}
\right], \nonumber\\
\frac{M_{\!_{+-}}}{L^2}
=& \int\!\frac{d^2 {\bf k}}{4\pi^2}\!
\left[ 
\frac{\hbar\omega}
{2E_{+} E_{-}}
\frac{E_{+}\xi_{-}+\xi_{+}E_{-}}
{\hbar^2\omega^2 - (E_{+}+E_{-})^2} \right], 
\nonumber
\end{align}
where $E_{\bf k} = \sqrt{\xi_{\bf k}^2+\vert\Delta\vert^2}$ is the quasiparticle
energy, $\xi_{\bf k} = \epsilon_{\bf k}  - \mu$ is the energy of the free
fermions of mass $m$ $(\epsilon_{\bf k} = \hbar^2\vert {\bf k} \vert^2/2m)$ with
respect to $\mu$, and $\vert \Delta \vert$ is the modulus of the order
parameter. We have used the shorthand notations $E_\pm = E_{\vk\pm \vq/2}$ and
$\xi_\pm = \xi_{\vk\pm \vq/2},\, M_{++} = M_{\theta\theta},\, M_{--} =
M_{\lambda\lambda}$, and $M_{+-} = M_{\theta \lambda}$. 

{\it Equation of state:} 
In Eq.~(\ref{eqn:action-gaussian}), the action $S_{\rm eff}$ is fully
characterized by $\vert \Delta \vert$ and $\mu$ or by the dimensionless
parameters $x = \mu/\vert\Delta\vert$ and $\vert\Delta\vert/\epsilon_F$, where
$\epsilon_F$ is the Fermi energy for a specified density $n = k_F^2/2\pi$, with
$k_F$ being the Fermi momentum. However, to study the evolution from the BCS to the
Bose limit, it is experimentally more relevant to relate $\vert \Delta \vert $
and $\mu$ to the 2D scattering length $a$ and the density $n$. The order
parameter is found from $\left[ \partial\Omega_{\rm sp}/\partial \vert \Delta
\vert \right]_{T,V}= 0$, where $\Omega_{\rm sp} = S_{\rm sp}/\beta$ is the
saddlepoint thermodynamic potential. Replacing the interaction strength $g$ in
favor of the (positive) two-body binding energy $\epsilon_{\rm b}$, using the
Lippmann-Schwinger relation $L^2/g =\sum_{\bf k} 1/(2\epsilon_{\bf k} +
\epsilon_b)$~\cite{Botelho2006}, one finds $\vert \Delta \vert =
\sqrt{\epsilon_b(2\mu+\epsilon_b)} \Theta(2\mu+\epsilon_b)$, which is explicitly
only a function of $\mu$ and $\epsilon_b$. The chemical potential can be found
by solving the saddlepoint number equation $n_{\rm sp} = -\left[
\partial\Omega_{\rm sp}/\partial\mu \right]_{T, V}/L^2$, while fixing the
density $n_{\rm sp}= n= k_F^2/2\pi$, resulting in $\mu_{\rm sp} =
\epsilon_F-\epsilon_b/2$, which substituted in the order parameter relation
leads to $\vert \Delta \vert_{\rm sp} = \sqrt{2\epsilon_F\epsilon_b}$. These
expressions are connected to the 2D scattering length $a$ via the relation
$\epsilon_b = 8\epsilon_F/\exp(2\gamma_\mathrm{E}+2\ln k_F a)$, with
$\gamma_\mathrm{E}\approx 0.577$ the Euler-Mascheroni
constant~\cite{Shlyapnikov2001}. However, as discussed below, this analysis
leads to the unphysical result of a constant sound velocity $c = v_F/\sqrt{2}$,
where $v_F$ is the Fermi velocity, over the entire BCS-to-Bose evolution,
because $\mu$ is inaccurately calculated.

\begin{figure}
\centering 
\includegraphics{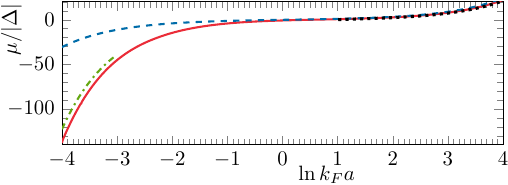}
\caption{
The ratio $x=\mu/\vert \Delta \vert$ for different approximations of the
equation of state: the saddlepoint approximation (dashed blue line) and
including Gaussian fluctuations (solid red line). The BCS and Bose limits are
shown as dotted black and dot-dashed green lines.
}
\label{fig:one}
\end{figure}

A more precise determination of $\mu$ for fixed density $n$ requires not only
$\Omega_{\rm sp}$, but also the Gaussian contribution $\Omega_{\rm g} = (k_B
T/2)\sum_q \ln \det [|\Delta| M(q)]$, found by integrating the bosonic fields in
the second term of Eq.~(\ref{eqn:action-gaussian}). The full number equation $n
= n_{\rm sp} + n_g$, including fluctuations of the order parameter at the
Gaussian level, is necessary to determine $\mu$. This is important both in
3D~\cite{Nozieres1985, SaDeMelo1993} and 2D~\cite{Botelho2006,HuLiu2015}. Here,
$n_g = -\left[ \partial\Omega_\mathrm{g}/\partial\mu \right]_{T, V}/L^2$ must be
always positive and $n (\mu) = k_F^2/2\pi$ must be solved numerically, leading
to a significant reduction of $\mu$ in the Bose regime~\footnote{Note that one
has to introduce convergence factors to regularize the Matsubara sum in the
thermodynamic potential $\Omega_\mathrm{g}$. This is most easily done in the
basis of the real and imaginary part of the pair field $\Phi$, instead of the
phase-modulus basis used here~\cite{SaDeMelo1997,HuLiu2015}.}, see
Fig.~\ref{fig:one}.

%

\begin{figure}
\centering 
\includegraphics{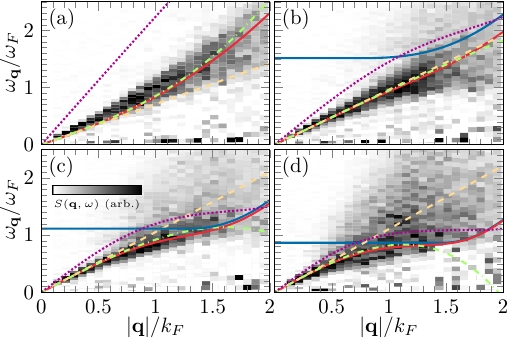}
\caption{
Frequency $\omega_{\bf q}$ vs. $\vert {\bf q} \vert$, using Fermi units
$\omega_F = \epsilon_F/\hbar$ and $k_F$, for different interactions. The solid
red line represents the numerical solution for $\omega_{\bf q}$, while the
analytical approximation is shown up to linear (dashed yellow line) and cubic
(dot-dashed green line) order in $\vert {\bf q} \vert$. The parameters for each
panel are: 
(a) $\ln k_F a=-0.1, \, (\mu/\vert \Delta \vert \simeq -1.0)$, 
(b) $\ln k_F a=0.7, \, (\mu/\vert \Delta \vert \simeq 0.2)$, 
(c) $\ln k_F a=1.1, \, (\mu/\vert \Delta \vert \simeq 0.7)$, and 
(d) $\ln k_F a=1.4, \, (\mu/\vert \Delta \vert\simeq 1.2)$,
using the Gaussian value of $\mu$. The dotted magenta lines represent
$\omega_{\bf q}^\mathrm{PO}$ using the saddlepoint $\mu$. The solid blue line
indicates the lower edge of the two-particle continuum $\epsilon_c(\vq)$ at the
Gaussian level. Our results are compared to dynamic structure factor experiments
(gray pixels) from Ref.~\cite{Lompe2021}, showing very good agreement.
}
\label{fig:two}
\end{figure}

{\it Collective Modes:}
As seen from Eq.~\eqref{eqn:action-gaussian}, the fluctuation matrix $M (q)$
acts as the inverse propagator of modulus and phase fluctuations. The collective
mode frequency $\omega_{\bf q}$ is found from the poles of $\left[ M
(q)\right]^{-1}$ or equivalently from $\det M(\vq,\omega_{\bf q}) =
0$~\cite{SaDeMelo1997}. As shown in the Supplemental Material, at $T=0$, these poles are
identical to the poles of the density-density response function, as probed by
experiments measuring the dynamical structure factor~\cite{Lompe2021}. A real
solution is found below the two-particle continuum $\epsilon_c(\vq) =
\min_\vk(E_++E_-)$, above which it is energetically more favorable to break
pairs. In a 3D Fermi gas, $\omega_{\bf q}$ can hit the two-particle continuum at
some finite value of ${\bf q}$ causing damping~\cite{Combescot2006}. However,
for a 2D Fermi gas, a real $\omega_{\bf q} < \epsilon_c ({\bf q})$ is found for
all values of ${\bf q}$, given that a two-body bound state always exists for a
2D contact potential~\cite{Landau1977}. This physics arises from $M_{++} =
M_{\theta\theta}$, which always diverges when $\hbar\omega \rightarrow
\epsilon_c$ making $\det M(\vq,\epsilon_c) = 0$ impossible, and thus there is no
damping of the mode at the Gaussian level (See Supplemental Material for an
in-depth discussion on damping). For arbitrary ${\bf q}$ there is no general
analytical solution for $\omega_{\bf q}$, but we obtain numerical results shown
as solid red lines in Fig.~\ref{fig:two}. We compare our results to the measured
spectrum from Ref.~\cite{Lompe2021}, and find that $\omega_{\bf q}$ follows
closely the maximum of the dynamical structure factor, without any fitting
parameters. Moreover, it can be seen that $\omega_{\bf q}$ avoids the
two-particle continuum $\epsilon_c ({\bf q})$ and that a solution exists for all
${\bf q}$.

\begin{figure}
\centering 
\includegraphics{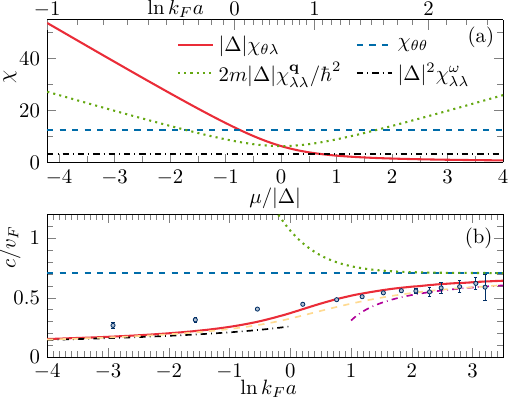}
\caption{
(a) Different $\chi$ coefficients vs. $\mu/\vert \Delta \vert$ or $\ln k_F a$ as
defined in Eq.~\eqref{propfun}. (b) sound velocity $c/v_F$ vs. $\ln k_F a$: the
dotted green line includes phase-only fluctuations with saddlepoint $\mu$,
diverging in the Bose limit; blue dashed line includes phase and modulus
fluctuations with saddlepoint $\mu$, giving always a constant value; solid red
line combines phase and modulus fluctuations with Gaussian $\mu$
selfconsistently. The dot-dashed magenta and black lines show the results in
the BCS and Bose limits, respectively. We compare our broken-U(1) sound velocity
$c$ with the experimental results of the isentropic sound velocity $u_S$ from
Ref.~\cite{Henning2020} (blue circles), and with $c$ from
Eq.~\eqref{eqn:speed-of-sound} and the Monte Carlo equation of state from
Ref.~\cite{Zhang2015} (dashed yellow line). See an in-depth discussion of the
conceptual difference between $c$ and $u_S$ in the Supplemental Material. 
}
\label{fig:three}
\end{figure}

Although numerical solutions are useful for comparison to recent
experiments~\cite{Henning2020,Lompe2021}, analytical insight is essential to
understand the underlying physics. To reveal the interplay between modulus and
phase fluctuations, we show next that their coupling increases dramatically as
the system evolves from the BCS to the Bose regime. We invert $M (q)$ and obtain
the propagators in Fourier space, which in the long wavelength limit become
\begin{align}
    &\begin{pmatrix}
        \langle \theta_{-q}\theta_q \rangle
            & \langle \theta_{-q}\lambda_q \rangle \\
        \langle \lambda_{-q}\theta_q \rangle
            & \langle \lambda_{-q}\lambda_q \rangle 
    \end{pmatrix}\label{propfun} \\
    &\quad=\frac{\vert \Delta \vert^2}{\hbar^2c^2q^2-\hbar^2\omega^2}\begin{pmatrix}
        \chi_{\theta\theta} & \ii \hbar\omega\, \chi_{\theta\lambda} \\
        -\ii \hbar\omega\, \chi_{\theta\lambda} 
        & \vq^2\, \chi^\vq_{\lambda\lambda} 
            - \hbar^2\omega^2\, \chi^\omega_{\lambda\lambda}
    \end{pmatrix}. \notag
\end{align}
The different $\chi$ coefficients defined in this equation are shown in
Fig.~\ref{fig:three}(a) as a function of interaction parametrized by the ratio
$x = \mu/\vert \Delta \vert$. They are explicitly given by
$\chi_{\theta\theta}=4\pi, \,
\chi_{\theta\lambda}=2\pi(-x+\sqrt{1+x^2})/|\Delta|, \,
\chi^\vq_{\lambda\lambda}=\hbar^2\pi\sqrt{1+x^2}/2m|\Delta|$, and
$\chi^\omega_{\lambda\lambda}=\pi/|\Delta|^2$. Most notably, the solid red line
in Fig.~\ref{fig:three}(a) shows $\chi_{\theta\lambda}$, which controls the
coupling between phase and modulus. Notice that $\chi_{\theta \lambda}$ is large
in the Bose regime $(\mu/\vert \Delta \vert \ll -1 )$, indicating that phase and
modulus are strongly mixed, while it is negligible in the BCS regime $(\mu/\vert
\Delta \vert \gg 1)$, showing that phase and modulus are essentially decoupled.
For $\omega > 0$, the pole of $\left[ M(\vert {\bf q} \vert,
\omega)\right]^{-1}$ occurs at $\omega_{\bf q} = c \vert {\bf q} \vert$, with 
\begin{equation}
2mc^2 = \mu + \sqrt{\mu^2+|\Delta|^2},
\label{eqn:speed-of-sound}
\end{equation}
where $c$ is the sound velocity associated with the broken U(1) symmetry. This
expression includes modulus and phase fluctuations of the order parameter, and
explicitly illustrates the dependence of $c$ on $\mu$. In
Fig.~\ref{fig:three}(b), we show the behavior of $c/v_F$ at three levels of
approximation: the dotted green line includes only phase fluctuations, the
dashed blue line includes phase and modulus fluctuations with the saddlepoint
value of $\mu$, while the solid red line includes phase and modulus fluctuations
with a selfconsistent Gaussian fluctuation value of $\mu$. It is clear that the
first two levels of approximation~\cite{Marini1998,Salasnich2013} give
completely unphysical results, in particular predicting the constant value
$c=v_F/\sqrt{2}$ for any coupling in the saddlepoint approximation, as quoted
in the literature~\cite{Marini1998, Salasnich2013}. However, we show here that
the inclusion of phase and modulus fluctuations with the correct $\mu$ leads to
the appropriate behavior of $c$ both in the Bose and BCS limits, giving results
that are surprisingly close to the experimentally measured isentropic sound
velocity~\cite{Henning2020}, that is typically a good estimate for Landau's
first sound.

The sound mode arising from the broken U(1) symmetry should not be confused with
Landau's first or second sound~\cite{Landau1941USSR}, as they are fundamentally
different. Our $T = 0$ microscopic collective mode can be directly observed in
measurements of the dynamic structure factor, and exists in the collisionless
regime. Conversely, first and second sound result from a phenomenological
decomposition of the superfluid into two components, and exist only in the
hydrodynamic regime~\cite{Martin1964, Martin1965}. In clarifying the difference
between the broken-U(1) and Landau's first sound, we show that Landau's first
sound velocity is always larger than $c$, see Supplemental Material. We also
note that the isentropic sound velocity~\cite{Henning2020} is not the same as
the sound velocity that can be extracted from the dynamical structure
factor~\cite{Lompe2021}.

Further insight is gained by studying the long wavelength limit of $S_{\rm eff}$
in Eq.~(\ref{eqn:action-gaussian}) by expanding the fluctuation matrix $M({\bf
q}, \omega)$ for small $\vq$ and $\omega$. Performing an inverse Fourier
transform back to real space, the second term in Eq.~(\ref{eqn:action-gaussian})
reduces to
\begin{equation}
S_g = \frac{1}{2}\int d^3r 
\left[ 
\rho_s (\nabla\theta)^2 + A (\hbar\partial_\tau\theta)^2
+\ii D \lambda \hbar\partial_\tau\theta + C \lambda^2
\right].
\label{eqn:action-local}
\end{equation}
The first two coefficients are the $T = 0$ superfluid density $2m\rho_s/\hbar^2
= n_\mathrm{sp}/2$ and the compressibility $ A =
(1+x/\sqrt{1+x^2})m/8\pi\hbar^2$, while $D=m \vert \Delta
\vert/\sqrt{1+x^2}/2\pi\hbar^2$ controls the phase-modulus coupling, and $C =
\vert\Delta\vert^2(1+x/\sqrt{1+x^2})m/2\pi\hbar^2$ describes the mass term for
the modulus fluctuations. Neglecting modulus fluctuations $(\lambda = 0)$ leads
to a sound velocity $c = \sqrt{\rho_s/A}/\hbar =
(\mu^2+|\Delta|^2)^{1/4}/\sqrt{m}$, shown as the dotted green line in
Fig.~\ref{fig:three}(b), which diverges in the Bose limit. However, when
$\lambda \ne 0$, the coupling between modulus and phase renormalizes $A$.
Integrating out $\lambda$ leads to a renormalized phase-only action with
unchanged superfluid density $\rho_{sR} = \rho_{s}$, but renormalized
compressibility $A_{R} = A+D^2/4C = m/4\pi\hbar^2$. This renormalization leads
to the corrected speed $c = \sqrt{\rho_{sR}/A_R}/\hbar$ given by
Eq.~(\ref{eqn:speed-of-sound}), as expected. This moreover leads to the
conclusion that the collective mode studied here is neither a Goldstone (pure
phase) nor Higgs (pure modulus) mode, because the mixing of phase and modulus
cannot be neglected.

{\it Change in concavity:}
To investigate the low-energy collective modes beyond linear dispersion, it is
necessary to expand $M_{\pm \pm}$ and $M_{+-}$ up to sixth order in $\omega$ and
$\vert {\bf q} \vert$. In this case, the condition $\det M ({\bf q}, \omega_{\bf
q}) = 0$ leads to
\begin{equation}
\omega_{\bf q} 
= c 
\vert{\bf q}\vert \left[ 
1 +\frac{\gamma}{8} 
\left( 
\frac{\hbar \vert {\bf q} \vert}{mc} 
\right)^2\!\!
+\frac{\eta}{16} 
\left( 
\frac{\hbar \vert {\bf q} \vert}{mc} 
\right)^4\!\!
+\mathcal{O}\!
\left( \frac{\hbar \vert {\bf q} \vert}{mc} \right)^6
\right]\! ,
\label{eqn:omega-quintic-order}
\end{equation}
where the coefficients of the cubic and quintic order corrections have
analytic expressions
\begin{align}
\gamma =&\, \frac{1}{24} 
\left(
1-4x^2-x \frac{7+4x^2}{\sqrt{1+x^2}} 
\right), \label{gamma}\\
\eta =&\, -\frac{365+2802x^2+2048x^4+160x^6}{23040(1+x^2)} \notag \\
&-x\frac{685+1813x^2+1064x^4+80x^6}{11520(1+x^2)^{3/2}}.
\label{eta}
\end{align}
In Fig.~\ref{fig:two}, we show $\omega_{\bf q}$ for various interaction regimes
and different levels of approximation. The solid red curves represent the full
numerical solutions, while the other curves represent the linear (dashed yellow)
and cubic (dot-dashed green) approximations in $\vert {\bf q} \vert$ with the
Gaussian corrected $\mu$. The dotted line represents the numerical phase-only
$(\lambda = 0)$ dispersion $\omega_{\bf q}^{PO}$ using the saddlepoint value of
$\mu$. Notice that $\omega_{\bf q}^{PO}$ severely overestimates the correct
$\omega_{\bf q}$ in the Bose regime, that is, $\omega_{\bf q}^{PO} \gg
\omega_{\bf q}$, while in the BCS limit $\omega_{\bf q}^{PO} \approx \omega_{\bf
q}$, because the modulus and phase fluctuations are nearly decoupled. The panels
in Fig.~\ref{fig:two} show that there is a change in curvature in the solid red
lines, also found in 3D Fermi gases~\cite{SaDeMelo1997, Combescot2006,
Kurkjian2016, Biss2021}, where the dispersion $\omega_{\bf q}$ is supersonic
$(\gamma > 0)$ in the Bose regime shown in panel (a), and subsonic $(\gamma < 0)$ in the BCS
regime shown in panel (d), where it bends downwards due to the
pairbreaking continuum. The coefficients of the nonlinear terms play a
significant role at larger momenta. While the cubic correction gives a good
approximation for the large momentum behavior in the Bose limit, as one moves
towards the BCS regime, progressively higher-order terms are needed to produce
the appropriate behavior.

\begin{figure}
\centering 
\includegraphics{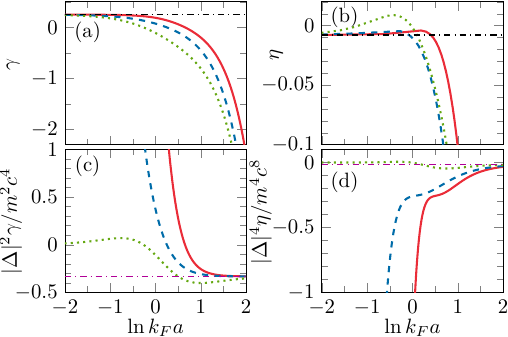}
\caption{
Parameters $\gamma$ and $\eta$ vs. $\ln k_F a$ at different levels of
approximation: Phase-only fluctuations with saddlepoint $\mu$ (dotted
green lines), phase and modulus fluctuations with saddlepoint $\mu$
(dashed blue lines) and with Gaussian corrected $\mu$ (solid red lines).
The dot-dashed black and magenta lines describe the Bose and BCS limits.
In (c) and (d), $\gamma$ and $\eta$ are scaled to reveal their
behavior in the BCS limit.
}
\label{fig:four}
\end{figure}

The coefficients $\gamma$ and $\eta$ are presented in Fig.~\ref{fig:four} as
function of $\ln k_F a$. In all panels, $\gamma$ and $\eta$ are evaluated at
different levels of approximation: the dotted green lines describe phase-only
fluctuations using the saddlepoint $\mu$, the dashed blue lines include modulus
and phase fluctuations using the saddlepoint $\mu$, and the solid red lines
include modulus and phase fluctuations using the Gaussian $\mu$. Panels (a)-(b)
((c)-(d)) show $\gamma$ and $\eta$ in units that elucidate their limiting
behavior in the Bose (BCS) regime given by the dot-dashed black (magenta) lines.

%

The concavity of $\omega_{\bf q}$ is controlled by $\gamma$, which changes from
$\gamma > 0$ (convex) to $\gamma < 0$ (concave) in the Bose and BCS regimes,
respectively. The parameter $\gamma$ changes sign at
$x=\sqrt{(2\sqrt{13}-7)/12}\approx 0.133$, corresponding to $\ln k_F a \simeq
0.65$ using the Gaussian $\mu$. In this case $(\gamma = 0)$, the first
correction to the linear spectrum is a quintic $(\vert {\bf q}\vert^5)$ term
controlled by $\eta$. Although the behavior of $\gamma$ and $\eta$ is similar to
the 3D results in the Bose limit~\cite{Kurkjian2016}, in the rest of the
crossover the 2D case is qualitatively different, where $\eta$ always stays
negative because of the strong level repulsion with the two-particle continuum,
due to the existence of two-body bound states for all interactions.

In the Bose limit $(x \ll -1)$, expanding the matrix elements of
$M({\bf q}, \omega)$ to order $(\Delta/|\mu|)^2$ and to lowest order in
$\hbar \vert{\bf q} \vert/\sqrt{2m|\mu|}$ leads to 
\begin{equation}
\hbar^2 \left(\omega_{\bf q}^{\rm B}\right)^2 
= 
\frac{\hbar^2q^2}{2m_B}\left( 
\frac{\hbar^2q^2}{2m_B} + 2m_B c^2_B
\right),
\label{eqn:bogoliubov-dispersion}
\end{equation}
where $c_B= \vert \Delta \vert/\sqrt{2m_B|\mu|}$ is the Bogoliubov speed of
sound, and $m_B = 2m$ is the boson mass.  Using the saddlepoint $\mu$ leads to
the incorrect value $c_B = v_F /\sqrt{2}$, while using the Gaussian corrected
$\mu$ leads to $c_B = v_F/\sqrt{8|\ln k_F a|}$ in the Bose limit. The
Bogoliubov-Popov interaction energy $2n_B V_B (0)  = 2m_B c^2 = E_F/\vert\ln k_F
a \vert$, with boson density $n_B = n/2$ leads to the boson-boson interaction
parameter $V_B (0) = (E_F/n)/\vert \ln k_F a \vert$. The values of $\gamma$ and
$\eta$ from Eq.~(\ref{eqn:bogoliubov-dispersion}) are equal to limiting results
obtained from Eqs.~(\ref{gamma}-\ref{eta}), that is, $\gamma \rightarrow 1/4$,
and $\eta \rightarrow -1/128$, as seen in Figs.~\ref{fig:four}(a)-(b).

In the BCS limit $(x \gg 1)$, the saddlepoint and Gaussian correction tend to
the same results, as fluctuations are less important. Rescaling energies by
$\vert \Delta \vert$, such that $\hbar\omega_{\bf q}/\vert \Delta \vert$ tends
to a universal function of $\hbar c \vert {\bf q} \vert/\vert \Delta \vert$,
leads to $(\vert \Delta \vert/mc^2)^2\gamma \rightarrow -1/3$ and $(\vert \Delta
\vert/mc^2)^4 \eta \rightarrow -1/72$, as revealed in
Figs.~\ref{fig:four}(c)-(d). This is a consequence of the two-particle continuum
pushing down the collective mode branch~\cite{Combescot2006, Kurkjian2016}. In
this case, the expansion in $\vert {\bf q} \vert$ is limited to $\hbar c \vert
{\bf q} \vert \leq 2 \vert \Delta \vert$ and $c\to v_F/\sqrt{2}$.

{\it Conclusions:} 
We analyzed low-energy collective modes of 2D $s$-wave Fermi superfluids from the BCS to
the Bose regime giving excellent results when compared to Bragg spectroscopy
experiments in 2D box potentials. We showed that quantum fluctuations in the
phase and modulus of the pairing order parameter are absolutely necessary to
give physically acceptable chemical potential and sound velocity. We presented
analytical results for the change in concavity of the collective mode dispersion
from convex to concave as contact interactions are changed from the BCS to the Bose
regime. The dispersion never hits the two-particle continuum threshold, due to
the existence of two-body bound states for arbitrarily small attractive $s$-wave
interactions in 2D.

\begin{acknowledgments}
{\it Acknowledgments:}
We thank the Belgian American Educational Foundation for support. We also thank
Henning Moritz and Lennart Sobirey for sharing their experimental data.
\end{acknowledgments}

\nocite{Popov1991,
Combescot2006,
Kurkjian2016,
Beliaev1958,
Landau-Khalatknikov1949,
Wyatt1992,
Wyatt2009,
Ketterle1998,
Walraven2005,
Kurkjian2016EPL,
Castin2019,
Kurkjian2020,
Griffin2010,
Feynman1954,
Landau1941USSR,
Landau1941PR,
Martin1964,
Martin1965}


\begin{thebibliography}{69}%
\makeatletter
\providecommand \@ifxundefined [1]{%
 \@ifx{#1\undefined}
}%
\providecommand \@ifnum [1]{%
 \ifnum #1\expandafter \@firstoftwo
 \else \expandafter \@secondoftwo
 \fi
}%
\providecommand \@ifx [1]{%
 \ifx #1\expandafter \@firstoftwo
 \else \expandafter \@secondoftwo
 \fi
}%
\providecommand \natexlab [1]{#1}%
\providecommand \enquote  [1]{``#1''}%
\providecommand \bibnamefont  [1]{#1}%
\providecommand \bibfnamefont [1]{#1}%
\providecommand \citenamefont [1]{#1}%
\providecommand \href@noop [0]{\@secondoftwo}%
\providecommand \href [0]{\begingroup \@sanitize@url \@href}%
\providecommand \@href[1]{\@@startlink{#1}\@@href}%
\providecommand \@@href[1]{\endgroup#1\@@endlink}%
\providecommand \@sanitize@url [0]{\catcode `\\12\catcode `\$12\catcode
  `\&12\catcode `\#12\catcode `\^12\catcode `\_12\catcode `\%12\relax}%
\providecommand \@@startlink[1]{}%
\providecommand \@@endlink[0]{}%
\providecommand \url  [0]{\begingroup\@sanitize@url \@url }%
\providecommand \@url [1]{\endgroup\@href {#1}{\urlprefix }}%
\providecommand \urlprefix  [0]{URL }%
\providecommand \Eprint [0]{\href }%
\providecommand \doibase [0]{https://doi.org/}%
\providecommand \selectlanguage [0]{\@gobble}%
\providecommand \bibinfo  [0]{\@secondoftwo}%
\providecommand \bibfield  [0]{\@secondoftwo}%
\providecommand \translation [1]{[#1]}%
\providecommand \BibitemOpen [0]{}%
\providecommand \bibitemStop [0]{}%
\providecommand \bibitemNoStop [0]{.\EOS\space}%
\providecommand \EOS [0]{\spacefactor3000\relax}%
\providecommand \BibitemShut  [1]{\csname bibitem#1\endcsname}%
\let\auto@bib@innerbib\@empty
\bibitem [{\citenamefont {Goldstone}(1961)}]{Goldstone1961}%
  \BibitemOpen
  \bibfield  {author} {\bibinfo {author} {\bibfnamefont {J.}~\bibnamefont
  {Goldstone}},\ }\bibfield  {title} {\bibinfo {title} {{Field theories with
  superconductor solutions}},\ }\href {https://doi.org/10.1007/BF02812722}
  {\bibfield  {journal} {\bibinfo  {journal} {Nuovo Cimento}\ }\textbf
  {\bibinfo {volume} {19}},\ \bibinfo {pages} {154} (\bibinfo {year}
  {1961})}\BibitemShut {NoStop}%
\bibitem [{\citenamefont {Higgs}(1966)}]{Higgs1966}%
  \BibitemOpen
  \bibfield  {author} {\bibinfo {author} {\bibfnamefont {P.~W.}\ \bibnamefont
  {Higgs}},\ }\bibfield  {title} {\bibinfo {title} {{Spontaneous symmetry
  breakdown without massless bosons}},\ }\href
  {https://doi.org/10.1103/PhysRev.145.1156} {\bibfield  {journal} {\bibinfo
  {journal} {Phys. Rev.}\ }\textbf {\bibinfo {volume} {145}},\ \bibinfo {pages}
  {1156} (\bibinfo {year} {1966})}\BibitemShut {NoStop}%
\bibitem [{\citenamefont {Englert}\ \emph {et~al.}(1966)\citenamefont
  {Englert}, \citenamefont {Brout},\ and\ \citenamefont {Thiry}}]{Englert1966}%
  \BibitemOpen
  \bibfield  {author} {\bibinfo {author} {\bibfnamefont {F.}~\bibnamefont
  {Englert}}, \bibinfo {author} {\bibfnamefont {R.}~\bibnamefont {Brout}},\
  and\ \bibinfo {author} {\bibfnamefont {M.~F.}\ \bibnamefont {Thiry}},\
  }\bibfield  {title} {\bibinfo {title} {{Vector mesons in presence of broken
  symmetry}},\ }\href {https://doi.org/10.1007/BF02752859} {\bibfield
  {journal} {\bibinfo  {journal} {Nuovo Cimento}\ }\textbf {\bibinfo {volume}
  {A43}},\ \bibinfo {pages} {244} (\bibinfo {year} {1966})}\BibitemShut
  {NoStop}%
\bibitem [{\citenamefont {Hong}\ \emph {et~al.}(2017)\citenamefont {Hong},
  \citenamefont {Matsumoto}, \citenamefont {Y.}, \citenamefont {Chen},
  \citenamefont {Gentile}, \citenamefont {Watson}, \citenamefont {Awwadi},
  \citenamefont {Turnbull}, \citenamefont {E.}, \citenamefont {Agrawal},
  \citenamefont {Toft-Petersen}, \citenamefont {Klemke}, \citenamefont
  {Coester}, \citenamefont {Schmidt},\ and\ \citenamefont
  {Tennant}}]{Tennant2017}%
  \BibitemOpen
  \bibfield  {author} {\bibinfo {author} {\bibfnamefont {T.}~\bibnamefont
  {Hong}}, \bibinfo {author} {\bibfnamefont {M.}~\bibnamefont {Matsumoto}},
  \bibinfo {author} {\bibfnamefont {Q.}~\bibnamefont {Y.}}, \bibinfo {author}
  {\bibfnamefont {W.}~\bibnamefont {Chen}}, \bibinfo {author} {\bibfnamefont
  {T.~R.}\ \bibnamefont {Gentile}}, \bibinfo {author} {\bibfnamefont
  {S.}~\bibnamefont {Watson}}, \bibinfo {author} {\bibfnamefont {F.~F.}\
  \bibnamefont {Awwadi}}, \bibinfo {author} {\bibfnamefont {M.~M.}\
  \bibnamefont {Turnbull}}, \bibinfo {author} {\bibfnamefont {D.~S.}\
  \bibnamefont {E.}}, \bibinfo {author} {\bibfnamefont {H.}~\bibnamefont
  {Agrawal}}, \bibinfo {author} {\bibfnamefont {R.}~\bibnamefont
  {Toft-Petersen}}, \bibinfo {author} {\bibfnamefont {B.}~\bibnamefont
  {Klemke}}, \bibinfo {author} {\bibfnamefont {K.}~\bibnamefont {Coester}},
  \bibinfo {author} {\bibfnamefont {K.~P.}\ \bibnamefont {Schmidt}},\ and\
  \bibinfo {author} {\bibfnamefont {D.~A.}\ \bibnamefont {Tennant}},\
  }\bibfield  {title} {\bibinfo {title} {{Higgs amplitude mode in a
  two-dimensional quantum antiferromagnet near the quantum critical point}},\
  }\href {https://doi.org/10.1038/NPHYS418} {\bibfield  {journal} {\bibinfo
  {journal} {Nature Physics}\ }\textbf {\bibinfo {volume} {13}},\ \bibinfo
  {pages} {638} (\bibinfo {year} {2017})}\BibitemShut {NoStop}%
\bibitem [{\citenamefont {Maiti}\ and\ \citenamefont
  {Hirschfeld}(2015)}]{Hisrchfeld2015}%
  \BibitemOpen
  \bibfield  {author} {\bibinfo {author} {\bibfnamefont {S.}~\bibnamefont
  {Maiti}}\ and\ \bibinfo {author} {\bibfnamefont {P.~J.}\ \bibnamefont
  {Hirschfeld}},\ }\bibfield  {title} {\bibinfo {title} {{Collective modes in
  superconductors with competing $s$- and $d$-wave interactions}},\ }\href
  {https://doi.org/10.1103/PhysRevB.92.094506} {\bibfield  {journal} {\bibinfo
  {journal} {Phys. Rev. B}\ }\textbf {\bibinfo {volume} {92}},\ \bibinfo
  {pages} {094506} (\bibinfo {year} {2015})}\BibitemShut {NoStop}%
\bibitem [{\citenamefont {Greco}\ \emph {et~al.}(2003)\citenamefont {Greco},
  \citenamefont {Colonna}, \citenamefont {Di~Toro},\ and\ \citenamefont
  {Matera}}]{Matera2003}%
  \BibitemOpen
  \bibfield  {author} {\bibinfo {author} {\bibfnamefont {V.}~\bibnamefont
  {Greco}}, \bibinfo {author} {\bibfnamefont {M.}~\bibnamefont {Colonna}},
  \bibinfo {author} {\bibfnamefont {M.}~\bibnamefont {Di~Toro}},\ and\ \bibinfo
  {author} {\bibfnamefont {F.}~\bibnamefont {Matera}},\ }\bibfield  {title}
  {\bibinfo {title} {{Collective modes of asymmetric nuclear matter in quantum
  hadrodynamics}},\ }\href {https://doi.org/10.1103/PhysRevC.67.015203}
  {\bibfield  {journal} {\bibinfo  {journal} {Phys. Rev. C}\ }\textbf {\bibinfo
  {volume} {67}},\ \bibinfo {pages} {015203} (\bibinfo {year}
  {2003})}\BibitemShut {NoStop}%
\bibitem [{\citenamefont {Yamamoto}\ \emph {et~al.}(2007)\citenamefont
  {Yamamoto}, \citenamefont {Tachibana}, \citenamefont {Hatsuda},\ and\
  \citenamefont {Baym}}]{Baym2007}%
  \BibitemOpen
  \bibfield  {author} {\bibinfo {author} {\bibfnamefont {N.}~\bibnamefont
  {Yamamoto}}, \bibinfo {author} {\bibfnamefont {M.}~\bibnamefont {Tachibana}},
  \bibinfo {author} {\bibfnamefont {T.}~\bibnamefont {Hatsuda}},\ and\ \bibinfo
  {author} {\bibfnamefont {G.}~\bibnamefont {Baym}},\ }\bibfield  {title}
  {\bibinfo {title} {{Phase structure, collective modes, and the axial anomaly
  in dense QCD}},\ }\href {https://doi.org/10.1103/PhysRevD.76.074001}
  {\bibfield  {journal} {\bibinfo  {journal} {Phys. Rev. D}\ }\textbf {\bibinfo
  {volume} {76}},\ \bibinfo {pages} {074001} (\bibinfo {year}
  {2007})}\BibitemShut {NoStop}%
\bibitem [{\citenamefont {Kobyakov}\ and\ \citenamefont
  {Pethick}(2013)}]{Pethick2013}%
  \BibitemOpen
  \bibfield  {author} {\bibinfo {author} {\bibfnamefont {D.}~\bibnamefont
  {Kobyakov}}\ and\ \bibinfo {author} {\bibfnamefont {C.~J.}\ \bibnamefont
  {Pethick}},\ }\bibfield  {title} {\bibinfo {title} {{Dynamics of the inner
  crust of neutron stars: Hydrodynamics, elasticity, and collective modes}},\
  }\href {https://doi.org/10.1103/PhysRevC.87.055803} {\bibfield  {journal}
  {\bibinfo  {journal} {Phys. Rev. C}\ }\textbf {\bibinfo {volume} {87}},\
  \bibinfo {pages} {055803} (\bibinfo {year} {2013})}\BibitemShut {NoStop}%
\bibitem [{\citenamefont {Mathur}\ and\ \citenamefont
  {Turton}(2014)}]{Turton2014}%
  \BibitemOpen
  \bibfield  {author} {\bibinfo {author} {\bibfnamefont {S.~D.}\ \bibnamefont
  {Mathur}}\ and\ \bibinfo {author} {\bibfnamefont {D.}~\bibnamefont
  {Turton}},\ }\bibfield  {title} {\bibinfo {title} {{Comments on black holes
  I: The possibility of complementarity}},\ }\href
  {https://doi.org/10.1007/JHEP01(2014)034} {\bibfield  {journal} {\bibinfo
  {journal} {J. High Energ. Phys.}\ }\textbf {\bibinfo {volume} {2014}},\
  \bibinfo {pages} {34}}\BibitemShut {NoStop}%
\bibitem [{\citenamefont {Singh}\ and\ \citenamefont
  {Mathey}(2021)}]{Mathey2021}%
  \BibitemOpen
  \bibfield  {author} {\bibinfo {author} {\bibfnamefont {V.~P.}\ \bibnamefont
  {Singh}}\ and\ \bibinfo {author} {\bibfnamefont {L.}~\bibnamefont {Mathey}},\
  }\bibfield  {title} {\bibinfo {title} {{Collective modes and superfluidity of
  a two-dimensional ultracold Bose gas}},\ }\href
  {https://doi.org/10.1103/PhysRevResearch.3.02311} {\bibfield  {journal}
  {\bibinfo  {journal} {Phys. Rev. Research}\ }\textbf {\bibinfo {volume}
  {3}},\ \bibinfo {pages} {023112} (\bibinfo {year} {2021})}\BibitemShut
  {NoStop}%
\bibitem [{\citenamefont {Choudhury}\ \emph {et~al.}(2020)\citenamefont
  {Choudhury}, \citenamefont {Islam}, \citenamefont {Hou}, \citenamefont
  {Aman}, \citenamefont {Killian},\ and\ \citenamefont
  {Hazzard}}]{Hazzard2020}%
  \BibitemOpen
  \bibfield  {author} {\bibinfo {author} {\bibfnamefont {S.}~\bibnamefont
  {Choudhury}}, \bibinfo {author} {\bibfnamefont {K.~R.}\ \bibnamefont
  {Islam}}, \bibinfo {author} {\bibfnamefont {Y.}~\bibnamefont {Hou}}, \bibinfo
  {author} {\bibfnamefont {J.~A.}\ \bibnamefont {Aman}}, \bibinfo {author}
  {\bibfnamefont {T.~C.}\ \bibnamefont {Killian}},\ and\ \bibinfo {author}
  {\bibfnamefont {K.~R.~A.}\ \bibnamefont {Hazzard}},\ }\bibfield  {title}
  {\bibinfo {title} {{Collective modes of ultracold fermionic
  alkaline-earth-metal gases with SU(N) symmetry}},\ }\href
  {https://doi.org/10.1103/PhysRevA.101.053612} {\bibfield  {journal} {\bibinfo
   {journal} {Phys. Rev. A}\ }\textbf {\bibinfo {volume} {101}},\ \bibinfo
  {pages} {053612} (\bibinfo {year} {2020})}\BibitemShut {NoStop}%
\bibitem [{\citenamefont {Chin}\ \emph {et~al.}(2010)\citenamefont {Chin},
  \citenamefont {Grimm}, \citenamefont {Julienne},\ and\ \citenamefont
  {Tiesinga}}]{Chin2010}%
  \BibitemOpen
  \bibfield  {author} {\bibinfo {author} {\bibfnamefont {C.}~\bibnamefont
  {Chin}}, \bibinfo {author} {\bibfnamefont {R.}~\bibnamefont {Grimm}},
  \bibinfo {author} {\bibfnamefont {P.}~\bibnamefont {Julienne}},\ and\
  \bibinfo {author} {\bibfnamefont {E.}~\bibnamefont {Tiesinga}},\ }\bibfield
  {title} {\bibinfo {title} {{Feshbach resonances in ultracold gases}},\ }\href
  {https://doi.org/10.1103/RevModPhys.82.1225} {\bibfield  {journal} {\bibinfo
  {journal} {Rev. Mod. Phys.}\ }\textbf {\bibinfo {volume} {82}},\ \bibinfo
  {pages} {1225} (\bibinfo {year} {2010})}\BibitemShut {NoStop}%
\bibitem [{\citenamefont {Ibarra-Garc\'{\i}a-Padilla}\ \emph
  {et~al.}(2020)\citenamefont {Ibarra-Garc\'{\i}a-Padilla}, \citenamefont
  {Mukherjee}, \citenamefont {Hulet}, \citenamefont {Hazzard}, \citenamefont
  {Paiva},\ and\ \citenamefont {Scalettar}}]{Hulet2020}%
  \BibitemOpen
  \bibfield  {author} {\bibinfo {author} {\bibfnamefont {E.}~\bibnamefont
  {Ibarra-Garc\'{\i}a-Padilla}}, \bibinfo {author} {\bibfnamefont
  {R.}~\bibnamefont {Mukherjee}}, \bibinfo {author} {\bibfnamefont {R.~G.}\
  \bibnamefont {Hulet}}, \bibinfo {author} {\bibfnamefont {K.~R.~A.}\
  \bibnamefont {Hazzard}}, \bibinfo {author} {\bibfnamefont {T.}~\bibnamefont
  {Paiva}},\ and\ \bibinfo {author} {\bibfnamefont {R.~T.}\ \bibnamefont
  {Scalettar}},\ }\bibfield  {title} {\bibinfo {title} {{Thermodynamics and
  magnetism in the two-dimensional to three-dimensional crossover of the
  Hubbard model}},\ }\href {https://doi.org/10.1103/PhysRevA.102.033340}
  {\bibfield  {journal} {\bibinfo  {journal} {Phys. Rev. A}\ }\textbf {\bibinfo
  {volume} {102}},\ \bibinfo {pages} {033340} (\bibinfo {year}
  {2020})}\BibitemShut {NoStop}%
\bibitem [{\citenamefont {Mermin}\ and\ \citenamefont
  {Wagner}(1966)}]{Mermin1966}%
  \BibitemOpen
  \bibfield  {author} {\bibinfo {author} {\bibfnamefont {N.~D.}\ \bibnamefont
  {Mermin}}\ and\ \bibinfo {author} {\bibfnamefont {H.}~\bibnamefont
  {Wagner}},\ }\bibfield  {title} {\bibinfo {title} {{Absence of ferromagnetism
  or antiferromagnetism in one- or two-dimensional isotropic Heisenberg
  models}},\ }\href {https://doi.org/10.1103/PhysRevLett.17.1133} {\bibfield
  {journal} {\bibinfo  {journal} {Phys. Rev. Lett.}\ }\textbf {\bibinfo
  {volume} {17}},\ \bibinfo {pages} {1133} (\bibinfo {year}
  {1966})}\BibitemShut {NoStop}%
\bibitem [{\citenamefont {Hohenberg}(1967)}]{Hohenberg1967}%
  \BibitemOpen
  \bibfield  {author} {\bibinfo {author} {\bibfnamefont {P.~C.}\ \bibnamefont
  {Hohenberg}},\ }\bibfield  {title} {\bibinfo {title} {{Existence of
  long-range order in one and two dimensions}},\ }\href
  {https://doi.org/10.1103/PhysRev.158.383} {\bibfield  {journal} {\bibinfo
  {journal} {Phys. Rev.}\ }\textbf {\bibinfo {volume} {158}},\ \bibinfo {pages}
  {383} (\bibinfo {year} {1967})}\BibitemShut {NoStop}%
\bibitem [{\citenamefont {Berezinskii}(1972)}]{Berezinskii1972}%
  \BibitemOpen
  \bibfield  {author} {\bibinfo {author} {\bibfnamefont {V.~L.}\ \bibnamefont
  {Berezinskii}},\ }\bibfield  {title} {\bibinfo {title} {{Destruction of
  long-range order in one-dimensional and two-dimensional systems possessing a
  continuous symmetry group. II. Quantum systems}},\ }\href
  {https://inspirehep.net/files/0f7b50c47ec26bed99a50ff199960259} {\bibfield
  {journal} {\bibinfo  {journal} {Sov. Phys. JETP}\ }\textbf {\bibinfo {volume}
  {34}},\ \bibinfo {pages} {610} (\bibinfo {year} {1972})}\BibitemShut
  {NoStop}%
\bibitem [{\citenamefont {Kosterlitz}\ and\ \citenamefont
  {Thouless}(1972)}]{Kosterlitz1972}%
  \BibitemOpen
  \bibfield  {author} {\bibinfo {author} {\bibfnamefont {J.~M.}\ \bibnamefont
  {Kosterlitz}}\ and\ \bibinfo {author} {\bibfnamefont {D.~J.}\ \bibnamefont
  {Thouless}},\ }\bibfield  {title} {\bibinfo {title} {{Long range order and
  metastability in two dimensional solids and superfluids.(Application of
  dislocation theory)}},\ }\href {https://doi.org/10.1088/0022-3719/5/11/002}
  {\bibfield  {journal} {\bibinfo  {journal} {Journal of Physics C: Solid State
  Physics}\ }\textbf {\bibinfo {volume} {5}},\ \bibinfo {pages} {L124}
  (\bibinfo {year} {1972})}\BibitemShut {NoStop}%
\bibitem [{\citenamefont {Keimer}\ \emph {et~al.}(2015)\citenamefont {Keimer},
  \citenamefont {Kivelson}, \citenamefont {Norman}, \citenamefont {Uchida},\
  and\ \citenamefont {Zaanen}}]{Keimer2015}%
  \BibitemOpen
  \bibfield  {author} {\bibinfo {author} {\bibfnamefont {B.}~\bibnamefont
  {Keimer}}, \bibinfo {author} {\bibfnamefont {S.~A.}\ \bibnamefont
  {Kivelson}}, \bibinfo {author} {\bibfnamefont {M.~R.}\ \bibnamefont
  {Norman}}, \bibinfo {author} {\bibfnamefont {S.}~\bibnamefont {Uchida}},\
  and\ \bibinfo {author} {\bibfnamefont {J.}~\bibnamefont {Zaanen}},\
  }\bibfield  {title} {\bibinfo {title} {{From quantum matter to
  high-temperature superconductivity in copper oxides}},\ }\href
  {https://doi.org/10.1038/nature14165} {\bibfield  {journal} {\bibinfo
  {journal} {Nature}\ }\textbf {\bibinfo {volume} {518}},\ \bibinfo {pages}
  {179} (\bibinfo {year} {2015})}\BibitemShut {NoStop}%
\bibitem [{\citenamefont {Ge}\ \emph {et~al.}(2015)\citenamefont {Ge},
  \citenamefont {Liu}, \citenamefont {Liu}, \citenamefont {Gao}, \citenamefont
  {Qian}, \citenamefont {Xue}, \citenamefont {Liu},\ and\ \citenamefont
  {Jia}}]{Ge2015}%
  \BibitemOpen
  \bibfield  {author} {\bibinfo {author} {\bibfnamefont {J.-F.}\ \bibnamefont
  {Ge}}, \bibinfo {author} {\bibfnamefont {Z.-L.}\ \bibnamefont {Liu}},
  \bibinfo {author} {\bibfnamefont {C.}~\bibnamefont {Liu}}, \bibinfo {author}
  {\bibfnamefont {C.-L.}\ \bibnamefont {Gao}}, \bibinfo {author} {\bibfnamefont
  {D.}~\bibnamefont {Qian}}, \bibinfo {author} {\bibfnamefont {Q.-K.}\
  \bibnamefont {Xue}}, \bibinfo {author} {\bibfnamefont {Y.}~\bibnamefont
  {Liu}},\ and\ \bibinfo {author} {\bibfnamefont {J.-F.}\ \bibnamefont {Jia}},\
  }\bibfield  {title} {\bibinfo {title} {{Superconductivity above 100 K in
  single-layer FeSe films on doped SrTiO$_3$}},\ }\href
  {https://doi.org/10.1038/nmat4153} {\bibfield  {journal} {\bibinfo  {journal}
  {Nature Materials}\ }\textbf {\bibinfo {volume} {14}},\ \bibinfo {pages}
  {285} (\bibinfo {year} {2015})}\BibitemShut {NoStop}%
\bibitem [{\citenamefont {Cao}\ \emph {et~al.}(2018)\citenamefont {Cao},
  \citenamefont {Fatemi}, \citenamefont {Fang}, \citenamefont {Watanabe},
  \citenamefont {Taniguchi}, \citenamefont {Kaxiras},\ and\ \citenamefont
  {Jarillo-Herrero}}]{Cao2018}%
  \BibitemOpen
  \bibfield  {author} {\bibinfo {author} {\bibfnamefont {Y.}~\bibnamefont
  {Cao}}, \bibinfo {author} {\bibfnamefont {V.}~\bibnamefont {Fatemi}},
  \bibinfo {author} {\bibfnamefont {S.}~\bibnamefont {Fang}}, \bibinfo {author}
  {\bibfnamefont {K.}~\bibnamefont {Watanabe}}, \bibinfo {author}
  {\bibfnamefont {T.}~\bibnamefont {Taniguchi}}, \bibinfo {author}
  {\bibfnamefont {E.}~\bibnamefont {Kaxiras}},\ and\ \bibinfo {author}
  {\bibfnamefont {P.}~\bibnamefont {Jarillo-Herrero}},\ }\bibfield  {title}
  {\bibinfo {title} {{Unconventional superconductivity in magic-angle graphene
  superlattices}},\ }\href {https://doi.org/10.1038/nature26160} {\bibfield
  {journal} {\bibinfo  {journal} {Nature}\ }\textbf {\bibinfo {volume} {556}},\
  \bibinfo {pages} {43} (\bibinfo {year} {2018})}\BibitemShut {NoStop}%
\bibitem [{\citenamefont {Yu}\ \emph {et~al.}(2019)\citenamefont {Yu},
  \citenamefont {Ma}, \citenamefont {Cai}, \citenamefont {Zhong}, \citenamefont
  {Ye}, \citenamefont {Shen}, \citenamefont {Gu}, \citenamefont {Chen},\ and\
  \citenamefont {Zhang}}]{Yu2019}%
  \BibitemOpen
  \bibfield  {author} {\bibinfo {author} {\bibfnamefont {Y.}~\bibnamefont
  {Yu}}, \bibinfo {author} {\bibfnamefont {L.}~\bibnamefont {Ma}}, \bibinfo
  {author} {\bibfnamefont {P.}~\bibnamefont {Cai}}, \bibinfo {author}
  {\bibfnamefont {R.}~\bibnamefont {Zhong}}, \bibinfo {author} {\bibfnamefont
  {C.}~\bibnamefont {Ye}}, \bibinfo {author} {\bibfnamefont {J.}~\bibnamefont
  {Shen}}, \bibinfo {author} {\bibfnamefont {G.~D.}\ \bibnamefont {Gu}},
  \bibinfo {author} {\bibfnamefont {X.~H.}\ \bibnamefont {Chen}},\ and\
  \bibinfo {author} {\bibfnamefont {Y.}~\bibnamefont {Zhang}},\ }\bibfield
  {title} {\bibinfo {title} {{High-temperature superconductivity in monolayer
  Bi$_2$Sr$_2$CaCu$_2$O$_{8+\delta}$}},\ }\href
  {https://doi.org/10.1038/s41586-019-1718-x} {\bibfield  {journal} {\bibinfo
  {journal} {Nature}\ }\textbf {\bibinfo {volume} {575}},\ \bibinfo {pages}
  {156} (\bibinfo {year} {2019})}\BibitemShut {NoStop}%
\bibitem [{\citenamefont {Makhalov}\ \emph {et~al.}(2014)\citenamefont
  {Makhalov}, \citenamefont {Martiyanov},\ and\ \citenamefont
  {Turlapov}}]{Turlapov2014}%
  \BibitemOpen
  \bibfield  {author} {\bibinfo {author} {\bibfnamefont {V.}~\bibnamefont
  {Makhalov}}, \bibinfo {author} {\bibfnamefont {K.}~\bibnamefont
  {Martiyanov}},\ and\ \bibinfo {author} {\bibfnamefont {A.}~\bibnamefont
  {Turlapov}},\ }\bibfield  {title} {\bibinfo {title} {{Ground-state pressure
  of quasi-2D Fermi and Bose gases}},\ }\href
  {https://doi.org/10.1103/PhysRevLett.112.045301} {\bibfield  {journal}
  {\bibinfo  {journal} {Phys. Rev. Lett.}\ }\textbf {\bibinfo {volume} {112}},\
  \bibinfo {pages} {045301} (\bibinfo {year} {2014})}\BibitemShut {NoStop}%
\bibitem [{\citenamefont {Fenech}\ \emph {et~al.}(2016)\citenamefont {Fenech},
  \citenamefont {Dyke}, \citenamefont {Peppler}, \citenamefont {Lingham},
  \citenamefont {Hoinka}, \citenamefont {Hu},\ and\ \citenamefont
  {Vale}}]{Vale2016}%
  \BibitemOpen
  \bibfield  {author} {\bibinfo {author} {\bibfnamefont {K.}~\bibnamefont
  {Fenech}}, \bibinfo {author} {\bibfnamefont {P.}~\bibnamefont {Dyke}},
  \bibinfo {author} {\bibfnamefont {T.}~\bibnamefont {Peppler}}, \bibinfo
  {author} {\bibfnamefont {M.~G.}\ \bibnamefont {Lingham}}, \bibinfo {author}
  {\bibfnamefont {S.}~\bibnamefont {Hoinka}}, \bibinfo {author} {\bibfnamefont
  {H.}~\bibnamefont {Hu}},\ and\ \bibinfo {author} {\bibfnamefont {C.~J.}\
  \bibnamefont {Vale}},\ }\bibfield  {title} {\bibinfo {title} {{Thermodynamics
  of an attractive 2D Fermi gas}},\ }\href
  {https://doi.org/10.1103/PhysRevLett.116.045302} {\bibfield  {journal}
  {\bibinfo  {journal} {Phys. Rev. Lett.}\ }\textbf {\bibinfo {volume} {116}},\
  \bibinfo {pages} {045302} (\bibinfo {year} {2016})}\BibitemShut {NoStop}%
\bibitem [{\citenamefont {Boettcher}\ \emph {et~al.}(2016)\citenamefont
  {Boettcher}, \citenamefont {Bayha}, \citenamefont {Kedar}, \citenamefont
  {Murthy}, \citenamefont {Neidig}, \citenamefont {Ries}, \citenamefont {Wenz},
  \citenamefont {Z\"urn}, \citenamefont {Jochim},\ and\ \citenamefont
  {Enss}}]{Enss2016}%
  \BibitemOpen
  \bibfield  {author} {\bibinfo {author} {\bibfnamefont {I.}~\bibnamefont
  {Boettcher}}, \bibinfo {author} {\bibfnamefont {L.}~\bibnamefont {Bayha}},
  \bibinfo {author} {\bibfnamefont {D.}~\bibnamefont {Kedar}}, \bibinfo
  {author} {\bibfnamefont {P.~A.}\ \bibnamefont {Murthy}}, \bibinfo {author}
  {\bibfnamefont {M.}~\bibnamefont {Neidig}}, \bibinfo {author} {\bibfnamefont
  {M.~G.}\ \bibnamefont {Ries}}, \bibinfo {author} {\bibfnamefont {A.~N.}\
  \bibnamefont {Wenz}}, \bibinfo {author} {\bibfnamefont {G.}~\bibnamefont
  {Z\"urn}}, \bibinfo {author} {\bibfnamefont {S.}~\bibnamefont {Jochim}},\
  and\ \bibinfo {author} {\bibfnamefont {T.}~\bibnamefont {Enss}},\ }\bibfield
  {title} {\bibinfo {title} {{Equation of state of ultracold fermions in the 2D
  BEC-BCS crossover region}},\ }\href
  {https://doi.org/10.1103/PhysRevLett.116.045303} {\bibfield  {journal}
  {\bibinfo  {journal} {Phys. Rev. Lett.}\ }\textbf {\bibinfo {volume} {116}},\
  \bibinfo {pages} {045303} (\bibinfo {year} {2016})}\BibitemShut {NoStop}%
\bibitem [{\citenamefont {Hadzibabic}\ \emph {et~al.}(2006)\citenamefont
  {Hadzibabic}, \citenamefont {Kr{\"u}ger}, \citenamefont {Cheneau},
  \citenamefont {Battelier},\ and\ \citenamefont {Dalibard}}]{Hadzibabic2006}%
  \BibitemOpen
  \bibfield  {author} {\bibinfo {author} {\bibfnamefont {Z.}~\bibnamefont
  {Hadzibabic}}, \bibinfo {author} {\bibfnamefont {P.}~\bibnamefont
  {Kr{\"u}ger}}, \bibinfo {author} {\bibfnamefont {M.}~\bibnamefont {Cheneau}},
  \bibinfo {author} {\bibfnamefont {B.}~\bibnamefont {Battelier}},\ and\
  \bibinfo {author} {\bibfnamefont {J.}~\bibnamefont {Dalibard}},\ }\bibfield
  {title} {\bibinfo {title} {{Berezinskii--Kosterlitz--Thouless crossover in a
  trapped atomic gas}},\ }\href {https://doi.org/10.1038/nature04851}
  {\bibfield  {journal} {\bibinfo  {journal} {Nature}\ }\textbf {\bibinfo
  {volume} {441}},\ \bibinfo {pages} {1118} (\bibinfo {year}
  {2006})}\BibitemShut {NoStop}%
\bibitem [{\citenamefont {Murthy}\ \emph {et~al.}(2015)\citenamefont {Murthy},
  \citenamefont {Boettcher}, \citenamefont {Bayha}, \citenamefont {Holzmann},
  \citenamefont {Kedar}, \citenamefont {Neidig}, \citenamefont {Ries},
  \citenamefont {Wenz}, \citenamefont {Z\"urn},\ and\ \citenamefont
  {Jochim}}]{Jochim2015}%
  \BibitemOpen
  \bibfield  {author} {\bibinfo {author} {\bibfnamefont {P.~A.}\ \bibnamefont
  {Murthy}}, \bibinfo {author} {\bibfnamefont {I.}~\bibnamefont {Boettcher}},
  \bibinfo {author} {\bibfnamefont {L.}~\bibnamefont {Bayha}}, \bibinfo
  {author} {\bibfnamefont {M.}~\bibnamefont {Holzmann}}, \bibinfo {author}
  {\bibfnamefont {D.}~\bibnamefont {Kedar}}, \bibinfo {author} {\bibfnamefont
  {M.}~\bibnamefont {Neidig}}, \bibinfo {author} {\bibfnamefont {M.~G.}\
  \bibnamefont {Ries}}, \bibinfo {author} {\bibfnamefont {A.~N.}\ \bibnamefont
  {Wenz}}, \bibinfo {author} {\bibfnamefont {G.}~\bibnamefont {Z\"urn}},\ and\
  \bibinfo {author} {\bibfnamefont {S.}~\bibnamefont {Jochim}},\ }\bibfield
  {title} {\bibinfo {title} {{Observation of the
  Berezinskii-Kosterlitz-Thouless phase transition in an ultracold Fermi
  gas}},\ }\href {https://doi.org/10.1103/PhysRevLett.115.010401} {\bibfield
  {journal} {\bibinfo  {journal} {Phys. Rev. Lett.}\ }\textbf {\bibinfo
  {volume} {115}},\ \bibinfo {pages} {010401} (\bibinfo {year}
  {2015})}\BibitemShut {NoStop}%
\bibitem [{\citenamefont {Ville}\ \emph {et~al.}(2018)\citenamefont {Ville},
  \citenamefont {Saint-Jalm}, \citenamefont {Le~Cerf}, \citenamefont
  {Aidelsburger}, \citenamefont {Nascimb\`ene}, \citenamefont {Dalibard},\ and\
  \citenamefont {Beugnon}}]{Dalibard2018}%
  \BibitemOpen
  \bibfield  {author} {\bibinfo {author} {\bibfnamefont {J.~L.}\ \bibnamefont
  {Ville}}, \bibinfo {author} {\bibfnamefont {R.}~\bibnamefont {Saint-Jalm}},
  \bibinfo {author} {\bibfnamefont {{\'{E}}.}~\bibnamefont {Le~Cerf}}, \bibinfo
  {author} {\bibfnamefont {M.}~\bibnamefont {Aidelsburger}}, \bibinfo {author}
  {\bibfnamefont {S.}~\bibnamefont {Nascimb\`ene}}, \bibinfo {author}
  {\bibfnamefont {J.}~\bibnamefont {Dalibard}},\ and\ \bibinfo {author}
  {\bibfnamefont {J.}~\bibnamefont {Beugnon}},\ }\bibfield  {title} {\bibinfo
  {title} {{Sound propagation in a uniform superfluid two-dimensional Bose
  gas}},\ }\href {https://doi.org/10.1103/PhysRevLett.121.145301} {\bibfield
  {journal} {\bibinfo  {journal} {Phys. Rev. Lett.}\ }\textbf {\bibinfo
  {volume} {121}},\ \bibinfo {pages} {145301} (\bibinfo {year}
  {2018})}\BibitemShut {NoStop}%
\bibitem [{\citenamefont {Bohlen}\ \emph {et~al.}(2020)\citenamefont {Bohlen},
  \citenamefont {Sobirey}, \citenamefont {Luick}, \citenamefont {Biss},
  \citenamefont {Enss}, \citenamefont {Lompe},\ and\ \citenamefont
  {Moritz}}]{Henning2020}%
  \BibitemOpen
  \bibfield  {author} {\bibinfo {author} {\bibfnamefont {M.}~\bibnamefont
  {Bohlen}}, \bibinfo {author} {\bibfnamefont {L.}~\bibnamefont {Sobirey}},
  \bibinfo {author} {\bibfnamefont {N.}~\bibnamefont {Luick}}, \bibinfo
  {author} {\bibfnamefont {H.}~\bibnamefont {Biss}}, \bibinfo {author}
  {\bibfnamefont {T.}~\bibnamefont {Enss}}, \bibinfo {author} {\bibfnamefont
  {T.}~\bibnamefont {Lompe}},\ and\ \bibinfo {author} {\bibfnamefont
  {H.}~\bibnamefont {Moritz}},\ }\bibfield  {title} {\bibinfo {title} {{Sound
  propagation and quantum-limited damping in a two-dimensional Fermi gas}},\
  }\href {https://doi.org/10.1103/PhysRevLett.124.240403} {\bibfield  {journal}
  {\bibinfo  {journal} {Phys. Rev. Lett.}\ }\textbf {\bibinfo {volume} {124}},\
  \bibinfo {pages} {240403} (\bibinfo {year} {2020})}\BibitemShut {NoStop}%
\bibitem [{\citenamefont {Christodoulou}\ \emph {et~al.}(2021)\citenamefont
  {Christodoulou}, \citenamefont {Ga{\l}ka}, \citenamefont {Dogra},
  \citenamefont {Lopes}, \citenamefont {Schmitt},\ and\ \citenamefont
  {Hadzibabic}}]{Christodoulou2021}%
  \BibitemOpen
  \bibfield  {author} {\bibinfo {author} {\bibfnamefont {P.}~\bibnamefont
  {Christodoulou}}, \bibinfo {author} {\bibfnamefont {M.}~\bibnamefont
  {Ga{\l}ka}}, \bibinfo {author} {\bibfnamefont {N.}~\bibnamefont {Dogra}},
  \bibinfo {author} {\bibfnamefont {R.}~\bibnamefont {Lopes}}, \bibinfo
  {author} {\bibfnamefont {J.}~\bibnamefont {Schmitt}},\ and\ \bibinfo {author}
  {\bibfnamefont {Z.}~\bibnamefont {Hadzibabic}},\ }\bibfield  {title}
  {\bibinfo {title} {{Observation of first and second sound in a BKT
  superfluid}},\ }\href {https://doi.org/10.1038/s41586-021-03537-9} {\bibfield
   {journal} {\bibinfo  {journal} {Nature}\ }\textbf {\bibinfo {volume}
  {594}},\ \bibinfo {pages} {191} (\bibinfo {year} {2021})}\BibitemShut
  {NoStop}%
\bibitem [{\citenamefont {Sobirey}\ \emph {et~al.}(2022)\citenamefont
  {Sobirey}, \citenamefont {Biss}, \citenamefont {Luick}, \citenamefont
  {Bohlen}, \citenamefont {Moritz},\ and\ \citenamefont {Lompe}}]{Lompe2021}%
  \BibitemOpen
  \bibfield  {author} {\bibinfo {author} {\bibfnamefont {L.}~\bibnamefont
  {Sobirey}}, \bibinfo {author} {\bibfnamefont {H.}~\bibnamefont {Biss}},
  \bibinfo {author} {\bibfnamefont {N.}~\bibnamefont {Luick}}, \bibinfo
  {author} {\bibfnamefont {M.}~\bibnamefont {Bohlen}}, \bibinfo {author}
  {\bibfnamefont {H.}~\bibnamefont {Moritz}},\ and\ \bibinfo {author}
  {\bibfnamefont {T.}~\bibnamefont {Lompe}},\ }\bibfield  {title} {\bibinfo
  {title} {{Observing the Influence of Reduced Dimensionality on Fermionic
  Superfluids}},\ }\href {https://doi.org/10.1103/PhysRevLett.129.083601}
  {\bibfield  {journal} {\bibinfo  {journal} {Phys. Rev. Lett.}\ }\textbf
  {\bibinfo {volume} {129}},\ \bibinfo {pages} {083601} (\bibinfo {year}
  {2022})}\BibitemShut {NoStop}%
\bibitem [{\citenamefont {Martiyanov}\ \emph {et~al.}(2010)\citenamefont
  {Martiyanov}, \citenamefont {Makhalov},\ and\ \citenamefont
  {Turlapov}}]{Turlapov2010}%
  \BibitemOpen
  \bibfield  {author} {\bibinfo {author} {\bibfnamefont {K.}~\bibnamefont
  {Martiyanov}}, \bibinfo {author} {\bibfnamefont {V.}~\bibnamefont
  {Makhalov}},\ and\ \bibinfo {author} {\bibfnamefont {A.}~\bibnamefont
  {Turlapov}},\ }\bibfield  {title} {\bibinfo {title} {{Observation of a
  two-dimensional Fermi gas of atoms}},\ }\href
  {https://doi.org/10.1103/PhysRevLett.105.030404} {\bibfield  {journal}
  {\bibinfo  {journal} {Phys. Rev. Lett.}\ }\textbf {\bibinfo {volume} {105}},\
  \bibinfo {pages} {030404} (\bibinfo {year} {2010})}\BibitemShut {NoStop}%
\bibitem [{\citenamefont {Feld}\ \emph {et~al.}(2011)\citenamefont {Feld},
  \citenamefont {Fr{\"o}hlich}, \citenamefont {Vogt}, \citenamefont
  {Koschorreck},\ and\ \citenamefont {K{\"o}hl}}]{Feld2011}%
  \BibitemOpen
  \bibfield  {author} {\bibinfo {author} {\bibfnamefont {M.}~\bibnamefont
  {Feld}}, \bibinfo {author} {\bibfnamefont {B.}~\bibnamefont {Fr{\"o}hlich}},
  \bibinfo {author} {\bibfnamefont {E.}~\bibnamefont {Vogt}}, \bibinfo {author}
  {\bibfnamefont {M.}~\bibnamefont {Koschorreck}},\ and\ \bibinfo {author}
  {\bibfnamefont {M.}~\bibnamefont {K{\"o}hl}},\ }\bibfield  {title} {\bibinfo
  {title} {{Observation of a pairing pseudogap in a two-dimensional Fermi
  gas}},\ }\href {https://doi.org/10.1038/nature10627} {\bibfield  {journal}
  {\bibinfo  {journal} {Nature}\ }\textbf {\bibinfo {volume} {480}},\ \bibinfo
  {pages} {75} (\bibinfo {year} {2011})}\BibitemShut {NoStop}%
\bibitem [{\citenamefont {Dyke}\ \emph {et~al.}(2011)\citenamefont {Dyke},
  \citenamefont {Kuhnle}, \citenamefont {Whitlock}, \citenamefont {Hu},
  \citenamefont {Mark}, \citenamefont {Hoinka}, \citenamefont {Lingham},
  \citenamefont {Hannaford},\ and\ \citenamefont {Vale}}]{Vale2011}%
  \BibitemOpen
  \bibfield  {author} {\bibinfo {author} {\bibfnamefont {P.}~\bibnamefont
  {Dyke}}, \bibinfo {author} {\bibfnamefont {E.~D.}\ \bibnamefont {Kuhnle}},
  \bibinfo {author} {\bibfnamefont {S.}~\bibnamefont {Whitlock}}, \bibinfo
  {author} {\bibfnamefont {H.}~\bibnamefont {Hu}}, \bibinfo {author}
  {\bibfnamefont {M.}~\bibnamefont {Mark}}, \bibinfo {author} {\bibfnamefont
  {S.}~\bibnamefont {Hoinka}}, \bibinfo {author} {\bibfnamefont
  {M.}~\bibnamefont {Lingham}}, \bibinfo {author} {\bibfnamefont
  {P.}~\bibnamefont {Hannaford}},\ and\ \bibinfo {author} {\bibfnamefont
  {C.~J.}\ \bibnamefont {Vale}},\ }\bibfield  {title} {\bibinfo {title}
  {{Crossover from 2D to 3D in a weakly interacting Fermi gas}},\ }\href
  {https://doi.org/10.1103/PhysRevLett.106.105304} {\bibfield  {journal}
  {\bibinfo  {journal} {Phys. Rev. Lett.}\ }\textbf {\bibinfo {volume} {106}},\
  \bibinfo {pages} {105304} (\bibinfo {year} {2011})}\BibitemShut {NoStop}%
\bibitem [{\citenamefont {Ong}\ \emph {et~al.}(2015)\citenamefont {Ong},
  \citenamefont {Cheng}, \citenamefont {Arakelyan},\ and\ \citenamefont
  {Thomas}}]{Thomas2015}%
  \BibitemOpen
  \bibfield  {author} {\bibinfo {author} {\bibfnamefont {W.}~\bibnamefont
  {Ong}}, \bibinfo {author} {\bibfnamefont {C.}~\bibnamefont {Cheng}}, \bibinfo
  {author} {\bibfnamefont {I.}~\bibnamefont {Arakelyan}},\ and\ \bibinfo
  {author} {\bibfnamefont {J.~E.}\ \bibnamefont {Thomas}},\ }\bibfield  {title}
  {\bibinfo {title} {{Spin-imbalanced quasi-two-dimensional Fermi gases}},\
  }\href {https://doi.org/10.1103/PhysRevLett.114.110403} {\bibfield  {journal}
  {\bibinfo  {journal} {Phys. Rev. Lett.}\ }\textbf {\bibinfo {volume} {114}},\
  \bibinfo {pages} {110403} (\bibinfo {year} {2015})}\BibitemShut {NoStop}%
\bibitem [{\citenamefont {Hueck}\ \emph {et~al.}(2018)\citenamefont {Hueck},
  \citenamefont {Luick}, \citenamefont {Sobirey}, \citenamefont {Siegl},
  \citenamefont {Lompe},\ and\ \citenamefont {Moritz}}]{Henning2018}%
  \BibitemOpen
  \bibfield  {author} {\bibinfo {author} {\bibfnamefont {K.}~\bibnamefont
  {Hueck}}, \bibinfo {author} {\bibfnamefont {N.}~\bibnamefont {Luick}},
  \bibinfo {author} {\bibfnamefont {L.}~\bibnamefont {Sobirey}}, \bibinfo
  {author} {\bibfnamefont {J.}~\bibnamefont {Siegl}}, \bibinfo {author}
  {\bibfnamefont {T.}~\bibnamefont {Lompe}},\ and\ \bibinfo {author}
  {\bibfnamefont {H.}~\bibnamefont {Moritz}},\ }\bibfield  {title} {\bibinfo
  {title} {{Two-dimensional homogeneous Fermi gases}},\ }\href
  {https://doi.org/10.1103/PhysRevLett.120.060402} {\bibfield  {journal}
  {\bibinfo  {journal} {Phys. Rev. Lett.}\ }\textbf {\bibinfo {volume} {120}},\
  \bibinfo {pages} {060402} (\bibinfo {year} {2018})}\BibitemShut {NoStop}%
\bibitem [{\citenamefont {Hoinka}\ \emph {et~al.}(2017)\citenamefont {Hoinka},
  \citenamefont {Dyke}, \citenamefont {Lingham}, \citenamefont {Kinnunen},
  \citenamefont {Bruun},\ and\ \citenamefont {Vale}}]{Hoinka2017}%
  \BibitemOpen
  \bibfield  {author} {\bibinfo {author} {\bibfnamefont {S.}~\bibnamefont
  {Hoinka}}, \bibinfo {author} {\bibfnamefont {P.}~\bibnamefont {Dyke}},
  \bibinfo {author} {\bibfnamefont {M.~G.}\ \bibnamefont {Lingham}}, \bibinfo
  {author} {\bibfnamefont {J.~J.}\ \bibnamefont {Kinnunen}}, \bibinfo {author}
  {\bibfnamefont {G.~M.}\ \bibnamefont {Bruun}},\ and\ \bibinfo {author}
  {\bibfnamefont {C.~J.}\ \bibnamefont {Vale}},\ }\bibfield  {title} {\bibinfo
  {title} {{Goldstone mode and pair-breaking excitations in atomic Fermi
  superfluids}},\ }\href {https://doi.org/10.1038/nphys4187} {\bibfield
  {journal} {\bibinfo  {journal} {Nature Physics}\ }\textbf {\bibinfo {volume}
  {13}},\ \bibinfo {pages} {943} (\bibinfo {year} {2017})}\BibitemShut
  {NoStop}%
\bibitem [{\citenamefont {Biss}\ \emph {et~al.}(2022)\citenamefont {Biss},
  \citenamefont {Sobirey}, \citenamefont {Luick}, \citenamefont {Bohlen},
  \citenamefont {Kinnunen}, \citenamefont {Bruun}, \citenamefont {Lompe},\ and\
  \citenamefont {Moritz}}]{Biss2021}%
  \BibitemOpen
  \bibfield  {author} {\bibinfo {author} {\bibfnamefont {H.}~\bibnamefont
  {Biss}}, \bibinfo {author} {\bibfnamefont {L.}~\bibnamefont {Sobirey}},
  \bibinfo {author} {\bibfnamefont {N.}~\bibnamefont {Luick}}, \bibinfo
  {author} {\bibfnamefont {M.}~\bibnamefont {Bohlen}}, \bibinfo {author}
  {\bibfnamefont {J.~J.}\ \bibnamefont {Kinnunen}}, \bibinfo {author}
  {\bibfnamefont {G.~M.}\ \bibnamefont {Bruun}}, \bibinfo {author}
  {\bibfnamefont {T.}~\bibnamefont {Lompe}},\ and\ \bibinfo {author}
  {\bibfnamefont {H.}~\bibnamefont {Moritz}},\ }\bibfield  {title} {\bibinfo
  {title} {{Excitation Spectrum and Superfluid Gap of an Ultracold Fermi
  Gas}},\ }\href {https://doi.org/10.1103/PhysRevLett.128.100401} {\bibfield
  {journal} {\bibinfo  {journal} {Phys. Rev. Lett.}\ }\textbf {\bibinfo
  {volume} {128}},\ \bibinfo {pages} {100401} (\bibinfo {year}
  {2022})}\BibitemShut {NoStop}%
\bibitem [{\citenamefont {Randeria}\ \emph {et~al.}(1990)\citenamefont
  {Randeria}, \citenamefont {Duan},\ and\ \citenamefont
  {Shieh}}]{Randeria1990}%
  \BibitemOpen
  \bibfield  {author} {\bibinfo {author} {\bibfnamefont {M.}~\bibnamefont
  {Randeria}}, \bibinfo {author} {\bibfnamefont {J.-M.}\ \bibnamefont {Duan}},\
  and\ \bibinfo {author} {\bibfnamefont {L.-Y.}\ \bibnamefont {Shieh}},\
  }\bibfield  {title} {\bibinfo {title} {{Superconductivity in a
  two-dimensional Fermi gas: Evolution from Cooper pairing to Bose
  condensation}},\ }\href {https://doi.org/10.1103/PhysRevB.41.327} {\bibfield
  {journal} {\bibinfo  {journal} {Phys. Rev. B}\ }\textbf {\bibinfo {volume}
  {41}},\ \bibinfo {pages} {327} (\bibinfo {year} {1990})}\BibitemShut
  {NoStop}%
\bibitem [{\citenamefont {Marini}\ \emph {et~al.}(1998)\citenamefont {Marini},
  \citenamefont {Pistolesi},\ and\ \citenamefont {Strinati}}]{Marini1998}%
  \BibitemOpen
  \bibfield  {author} {\bibinfo {author} {\bibfnamefont {M.}~\bibnamefont
  {Marini}}, \bibinfo {author} {\bibfnamefont {F.}~\bibnamefont {Pistolesi}},\
  and\ \bibinfo {author} {\bibfnamefont {G.}~\bibnamefont {Strinati}},\
  }\bibfield  {title} {\bibinfo {title} {{Evolution from BCS superconductivity
  to Bose condensation: analytic results for the crossover in three
  dimensions}},\ }\href {https://doi.org/10.1007/s100510050165} {\bibfield
  {journal} {\bibinfo  {journal} {The European Physical Journal B - Condensed
  Matter and Complex Systems}\ }\textbf {\bibinfo {volume} {1}},\ \bibinfo
  {pages} {151} (\bibinfo {year} {1998})}\BibitemShut {NoStop}%
\bibitem [{\citenamefont {Salasnich}\ \emph {et~al.}(2013)\citenamefont
  {Salasnich}, \citenamefont {Marchetti},\ and\ \citenamefont
  {Toigo}}]{Salasnich2013}%
  \BibitemOpen
  \bibfield  {author} {\bibinfo {author} {\bibfnamefont {L.}~\bibnamefont
  {Salasnich}}, \bibinfo {author} {\bibfnamefont {P.~A.}\ \bibnamefont
  {Marchetti}},\ and\ \bibinfo {author} {\bibfnamefont {F.}~\bibnamefont
  {Toigo}},\ }\bibfield  {title} {\bibinfo {title} {{Superfluidity, sound
  velocity, and quasicondensation in the two-dimensional BCS-BEC crossover}},\
  }\href {https://doi.org/10.1103/PhysRevA.88.053612} {\bibfield  {journal}
  {\bibinfo  {journal} {Phys. Rev. A}\ }\textbf {\bibinfo {volume} {88}},\
  \bibinfo {pages} {053612} (\bibinfo {year} {2013})}\BibitemShut {NoStop}%
\bibitem [{\citenamefont {Lumbeeck}\ \emph {et~al.}(2020)\citenamefont
  {Lumbeeck}, \citenamefont {Tempere},\ and\ \citenamefont
  {Klimin}}]{Lumbeeck2020}%
  \BibitemOpen
  \bibfield  {author} {\bibinfo {author} {\bibfnamefont {L.-P.}\ \bibnamefont
  {Lumbeeck}}, \bibinfo {author} {\bibfnamefont {J.}~\bibnamefont {Tempere}},\
  and\ \bibinfo {author} {\bibfnamefont {S.}~\bibnamefont {Klimin}},\
  }\bibfield  {title} {\bibinfo {title} {{Dispersion and damping of phononic
  excitations in Fermi superfluid gases in 2D}},\ }\href
  {https://doi.org/10.3390/condmat5010013} {\bibfield  {journal} {\bibinfo
  {journal} {Condensed Matter}\ }\textbf {\bibinfo {volume} {5}},\ \bibinfo
  {pages} {13} (\bibinfo {year} {2020})}\BibitemShut {NoStop}%
\bibitem [{\citenamefont {He}\ \emph {et~al.}(2015)\citenamefont {He},
  \citenamefont {L\"u}, \citenamefont {Cao}, \citenamefont {Hu},\ and\
  \citenamefont {Liu}}]{HuLiu2015}%
  \BibitemOpen
  \bibfield  {author} {\bibinfo {author} {\bibfnamefont {L.}~\bibnamefont
  {He}}, \bibinfo {author} {\bibfnamefont {H.}~\bibnamefont {L\"u}}, \bibinfo
  {author} {\bibfnamefont {G.}~\bibnamefont {Cao}}, \bibinfo {author}
  {\bibfnamefont {H.}~\bibnamefont {Hu}},\ and\ \bibinfo {author}
  {\bibfnamefont {X.-J.}\ \bibnamefont {Liu}},\ }\bibfield  {title} {\bibinfo
  {title} {{Quantum fluctuations in the BCS-BEC crossover of two-dimensional
  Fermi gases}},\ }\href {https://doi.org/10.1103/PhysRevA.92.023620}
  {\bibfield  {journal} {\bibinfo  {journal} {Phys. Rev. A}\ }\textbf {\bibinfo
  {volume} {92}},\ \bibinfo {pages} {023620} (\bibinfo {year}
  {2015})}\BibitemShut {NoStop}%
\bibitem [{\citenamefont {Bighin}\ and\ \citenamefont
  {Salasnich}(2016)}]{Salasnich2016}%
  \BibitemOpen
  \bibfield  {author} {\bibinfo {author} {\bibfnamefont {G.}~\bibnamefont
  {Bighin}}\ and\ \bibinfo {author} {\bibfnamefont {L.}~\bibnamefont
  {Salasnich}},\ }\bibfield  {title} {\bibinfo {title} {{Finite-temperature
  quantum fluctuations in two-dimensional Fermi superfluids}},\ }\href
  {https://doi.org/10.1103/PhysRevB.93.014519} {\bibfield  {journal} {\bibinfo
  {journal} {Phys. Rev. B}\ }\textbf {\bibinfo {volume} {93}},\ \bibinfo
  {pages} {014519} (\bibinfo {year} {2016})}\BibitemShut {NoStop}%
\bibitem [{\citenamefont {Popov}(1991)}]{Popov1991}%
  \BibitemOpen
  \bibfield  {author} {\bibinfo {author} {\bibfnamefont {V.~N.}\ \bibnamefont
  {Popov}},\ }\bibfield  {title} {\bibinfo {title} {Functional integrals and
  collective excitations},\ }\href@noop {} {\bibfield  {journal} {\bibinfo
  {journal} {Cambridge University Press}\ } (\bibinfo {year}
  {1991})}\BibitemShut {NoStop}%
\bibitem [{\citenamefont {Combescot}\ \emph {et~al.}(2006)\citenamefont
  {Combescot}, \citenamefont {Kagan},\ and\ \citenamefont
  {Stringari}}]{Combescot2006}%
  \BibitemOpen
  \bibfield  {author} {\bibinfo {author} {\bibfnamefont {R.}~\bibnamefont
  {Combescot}}, \bibinfo {author} {\bibfnamefont {M.~Y.}\ \bibnamefont
  {Kagan}},\ and\ \bibinfo {author} {\bibfnamefont {S.}~\bibnamefont
  {Stringari}},\ }\bibfield  {title} {\bibinfo {title} {{Collective mode of
  homogeneous superfluid Fermi gases in the BEC-BCS crossover}},\ }\href
  {https://doi.org/10.1103/PhysRevA.74.042717} {\bibfield  {journal} {\bibinfo
  {journal} {Phys. Rev. A}\ }\textbf {\bibinfo {volume} {74}},\ \bibinfo
  {pages} {042717} (\bibinfo {year} {2006})}\BibitemShut {NoStop}%
\bibitem [{\citenamefont {Kurkjian}\ \emph
  {et~al.}(2016{\natexlab{a}})\citenamefont {Kurkjian}, \citenamefont
  {Castin},\ and\ \citenamefont {Sinatra}}]{Kurkjian2016}%
  \BibitemOpen
  \bibfield  {author} {\bibinfo {author} {\bibfnamefont {H.}~\bibnamefont
  {Kurkjian}}, \bibinfo {author} {\bibfnamefont {Y.}~\bibnamefont {Castin}},\
  and\ \bibinfo {author} {\bibfnamefont {A.}~\bibnamefont {Sinatra}},\
  }\bibfield  {title} {\bibinfo {title} {{Concavity of the collective
  excitation branch of a Fermi gas in the BEC-BCS crossover}},\ }\href
  {https://doi.org/10.1103/PhysRevA.93.013623} {\bibfield  {journal} {\bibinfo
  {journal} {Phys. Rev. A}\ }\textbf {\bibinfo {volume} {93}},\ \bibinfo
  {pages} {013623} (\bibinfo {year} {2016}{\natexlab{a}})}\BibitemShut
  {NoStop}%
\bibitem [{\citenamefont {Engelbrecht}\ \emph {et~al.}(1997)\citenamefont
  {Engelbrecht}, \citenamefont {Randeria},\ and\ \citenamefont {S\'a~de
  Melo}}]{SaDeMelo1997}%
  \BibitemOpen
  \bibfield  {author} {\bibinfo {author} {\bibfnamefont {J.~R.}\ \bibnamefont
  {Engelbrecht}}, \bibinfo {author} {\bibfnamefont {M.}~\bibnamefont
  {Randeria}},\ and\ \bibinfo {author} {\bibfnamefont {C.~A.~R.}\ \bibnamefont
  {S\'a~de Melo}},\ }\bibfield  {title} {\bibinfo {title} {{BCS to Bose
  crossover: Broken-symmetry state}},\ }\href
  {https://doi.org/10.1103/PhysRevB.55.15153} {\bibfield  {journal} {\bibinfo
  {journal} {Phys. Rev. B}\ }\textbf {\bibinfo {volume} {55}},\ \bibinfo
  {pages} {15153} (\bibinfo {year} {1997})}\BibitemShut {NoStop}%
\bibitem [{\citenamefont {Botelho}\ and\ \citenamefont {S\'a~de
  Melo}(2006)}]{Botelho2006}%
  \BibitemOpen
  \bibfield  {author} {\bibinfo {author} {\bibfnamefont {S.~S.}\ \bibnamefont
  {Botelho}}\ and\ \bibinfo {author} {\bibfnamefont {C.~A.~R.}\ \bibnamefont
  {S\'a~de Melo}},\ }\bibfield  {title} {\bibinfo {title} {{Vortex-antivortex
  lattice in ultracold fermionic gases}},\ }\href
  {https://doi.org/10.1103/PhysRevLett.96.040404} {\bibfield  {journal}
  {\bibinfo  {journal} {Phys. Rev. Lett.}\ }\textbf {\bibinfo {volume} {96}},\
  \bibinfo {pages} {040404} (\bibinfo {year} {2006})}\BibitemShut {NoStop}%
\bibitem [{\citenamefont {Petrov}\ and\ \citenamefont
  {Shlyapnikov}(2001)}]{Shlyapnikov2001}%
  \BibitemOpen
  \bibfield  {author} {\bibinfo {author} {\bibfnamefont {D.~S.}\ \bibnamefont
  {Petrov}}\ and\ \bibinfo {author} {\bibfnamefont {G.~V.}\ \bibnamefont
  {Shlyapnikov}},\ }\bibfield  {title} {\bibinfo {title} {{Interatomic
  collisions in a tightly confined Bose gas}},\ }\href
  {https://doi.org/10.1103/PhysRevA.64.012706} {\bibfield  {journal} {\bibinfo
  {journal} {Phys. Rev. A}\ }\textbf {\bibinfo {volume} {64}},\ \bibinfo
  {pages} {012706} (\bibinfo {year} {2001})}\BibitemShut {NoStop}%
\bibitem [{\citenamefont {Nozi\`eres}\ and\ \citenamefont
  {Schmitt-Rink}(1985)}]{Nozieres1985}%
  \BibitemOpen
  \bibfield  {author} {\bibinfo {author} {\bibfnamefont {P.}~\bibnamefont
  {Nozi\`eres}}\ and\ \bibinfo {author} {\bibfnamefont {S.}~\bibnamefont
  {Schmitt-Rink}},\ }\bibfield  {title} {\bibinfo {title} {{Bose condensation
  in an attractive fermion gas: From weak to strong coupling
  superconductivity}},\ }\href {https://doi.org/10.1007/BF00683774} {\bibfield
  {journal} {\bibinfo  {journal} {J. Low Temp. Phys.}\ }\textbf {\bibinfo
  {volume} {59}},\ \bibinfo {pages} {195} (\bibinfo {year} {1985})}\BibitemShut
  {NoStop}%
\bibitem [{\citenamefont {S\'a~de Melo}\ \emph {et~al.}(1993)\citenamefont
  {S\'a~de Melo}, \citenamefont {Randeria},\ and\ \citenamefont
  {Engelbrecht}}]{SaDeMelo1993}%
  \BibitemOpen
  \bibfield  {author} {\bibinfo {author} {\bibfnamefont {C.~A.~R.}\
  \bibnamefont {S\'a~de Melo}}, \bibinfo {author} {\bibfnamefont
  {M.}~\bibnamefont {Randeria}},\ and\ \bibinfo {author} {\bibfnamefont
  {J.~R.}\ \bibnamefont {Engelbrecht}},\ }\bibfield  {title} {\bibinfo {title}
  {{Crossover from BCS to Bose superconductivity: Transition temperature and
  time-dependent Ginzburg-Landau theory}},\ }\href
  {https://doi.org/10.1103/PhysRevLett.71.3202} {\bibfield  {journal} {\bibinfo
   {journal} {Phys. Rev. Lett.}\ }\textbf {\bibinfo {volume} {71}},\ \bibinfo
  {pages} {3202} (\bibinfo {year} {1993})}\BibitemShut {NoStop}%
\bibitem [{Note1()}]{Note1}%
  \BibitemOpen
  \bibinfo {note} {Note that one has to introduce convergence factors to
  regularize the Matsubara sum in the thermodynamic potential $\Omega _\protect
  \mathrm {g}$. This is most easily done in the basis of the real and imaginary
  part of the pair field $\Phi $, instead of the phase-modulus basis used
  here~\cite {SaDeMelo1997,HuLiu2015}.}\BibitemShut {Stop}%
\bibitem [{\citenamefont {Landau}\ and\ \citenamefont
  {Lifshitz}(1977)}]{Landau1977}%
  \BibitemOpen
  \bibfield  {author} {\bibinfo {author} {\bibfnamefont {L.~D.}\ \bibnamefont
  {Landau}}\ and\ \bibinfo {author} {\bibfnamefont {E.~M.}\ \bibnamefont
  {Lifshitz}},\ }\bibfield  {title} {\bibinfo {title} {{Quantum Mechanics -
  $3^{\rm rd}$ edition}},\ }\href@noop {} {\bibfield  {journal} {\bibinfo
  {journal} {Pergamon Press}\ ,\ \bibinfo {pages} {63}} (\bibinfo {year}
  {1977})}\BibitemShut {NoStop}%
\bibitem [{\citenamefont {Shi}\ \emph {et~al.}(2015)\citenamefont {Shi},
  \citenamefont {Chiesa},\ and\ \citenamefont {Zhang}}]{Zhang2015}%
  \BibitemOpen
  \bibfield  {author} {\bibinfo {author} {\bibfnamefont {H.}~\bibnamefont
  {Shi}}, \bibinfo {author} {\bibfnamefont {S.}~\bibnamefont {Chiesa}},\ and\
  \bibinfo {author} {\bibfnamefont {S.}~\bibnamefont {Zhang}},\ }\bibfield
  {title} {\bibinfo {title} {{Ground-state properties of strongly interacting
  Fermi gases in two dimensions}},\ }\href
  {https://doi.org/10.1103/PhysRevA.92.033603} {\bibfield  {journal} {\bibinfo
  {journal} {Phys. Rev. A}\ }\textbf {\bibinfo {volume} {92}},\ \bibinfo
  {pages} {033603} (\bibinfo {year} {2015})}\BibitemShut {NoStop}%
\bibitem [{\citenamefont {Landau}(1941{\natexlab{a}})}]{Landau1941USSR}%
  \BibitemOpen
  \bibfield  {author} {\bibinfo {author} {\bibfnamefont {L.}~\bibnamefont
  {Landau}},\ }\bibfield  {title} {\bibinfo {title} {{Theory of the
  Superfluidity of Helium II}},\ }\href@noop {} {\bibfield  {journal} {\bibinfo
   {journal} {J. Phys. U.S.S.R.}\ }\textbf {\bibinfo {volume} {5}},\ \bibinfo
  {pages} {71} (\bibinfo {year} {1941}{\natexlab{a}})}\BibitemShut {NoStop}%
\bibitem [{\citenamefont {Hohenberg}\ and\ \citenamefont
  {Martin}(1964)}]{Martin1964}%
  \BibitemOpen
  \bibfield  {author} {\bibinfo {author} {\bibfnamefont {P.~C.}\ \bibnamefont
  {Hohenberg}}\ and\ \bibinfo {author} {\bibfnamefont {P.~C.}\ \bibnamefont
  {Martin}},\ }\bibfield  {title} {\bibinfo {title} {{Superfluid Dynamics in
  the Hydrodynamic $(\omega\tau \ll 1)$ and Collisionless $(\omega\tau \gg 1)$
  Domains}},\ }\href {https://doi.org/10.1103/PhysRevLett.12.69} {\bibfield
  {journal} {\bibinfo  {journal} {Phys. Rev. Lett.}\ }\textbf {\bibinfo
  {volume} {12}},\ \bibinfo {pages} {69} (\bibinfo {year} {1964})}\BibitemShut
  {NoStop}%
\bibitem [{\citenamefont {Hohenberg}\ and\ \citenamefont
  {Martin}(1965)}]{Martin1965}%
  \BibitemOpen
  \bibfield  {author} {\bibinfo {author} {\bibfnamefont {P.~C.}\ \bibnamefont
  {Hohenberg}}\ and\ \bibinfo {author} {\bibfnamefont {P.~C.}\ \bibnamefont
  {Martin}},\ }\bibfield  {title} {\bibinfo {title} {{Microscopic Theory of
  Superfluid Helium}},\ }\href {https://doi.org/10.1016/0003-4916(65)90280-0}
  {\bibfield  {journal} {\bibinfo  {journal} {Ann. Phys., NY}\ }\textbf
  {\bibinfo {volume} {34}},\ \bibinfo {pages} {291} (\bibinfo {year}
  {1965})}\BibitemShut {NoStop}%
\bibitem [{\citenamefont {Beliaev}(1958)}]{Beliaev1958}%
  \BibitemOpen
  \bibfield  {author} {\bibinfo {author} {\bibfnamefont {S.~T.}\ \bibnamefont
  {Beliaev}},\ }\bibfield  {title} {\bibinfo {title} {{Energy spectrum of
  non-ideal Bose gas}},\ }\href@noop {} {\bibfield  {journal} {\bibinfo
  {journal} {Sov. Phys. JETP}\ }\textbf {\bibinfo {volume} {34}},\ \bibinfo
  {pages} {299} (\bibinfo {year} {1958})}\BibitemShut {NoStop}%
\bibitem [{\citenamefont {Landau}\ and\ \citenamefont
  {Khalatnikov}(1949)}]{Landau-Khalatknikov1949}%
  \BibitemOpen
  \bibfield  {author} {\bibinfo {author} {\bibfnamefont {L.}~\bibnamefont
  {Landau}}\ and\ \bibinfo {author} {\bibfnamefont {I.}~\bibnamefont
  {Khalatnikov}},\ }\bibfield  {title} {\bibinfo {title} {{The theory of the
  viscosity of Helium II. I. Collisions of elementary excitations in Helium
  II}},\ }\href@noop {} {\bibfield  {journal} {\bibinfo  {journal} {Zh. Eksp.
  Teor. Fiz.}\ }\textbf {\bibinfo {volume} {19}},\ \bibinfo {pages} {637}
  (\bibinfo {year} {1949})}\BibitemShut {NoStop}%
\bibitem [{\citenamefont {Tucker}\ and\ \citenamefont
  {Wyatt}(1992)}]{Wyatt1992}%
  \BibitemOpen
  \bibfield  {author} {\bibinfo {author} {\bibfnamefont {M.~A.~H.}\
  \bibnamefont {Tucker}}\ and\ \bibinfo {author} {\bibfnamefont {A.~F.~G.}\
  \bibnamefont {Wyatt}},\ }\bibfield  {title} {\bibinfo {title} {{Four-phonon
  scattering in superfluid $^4$He}},\ }\href
  {https://iopscience.iop.org/article/10.1088/0953-8984/4/38/008/pdf}
  {\bibfield  {journal} {\bibinfo  {journal} {J. Phys.: Condens. Matter}\
  }\textbf {\bibinfo {volume} {4}},\ \bibinfo {pages} {7745} (\bibinfo {year}
  {1992})}\BibitemShut {NoStop}%
\bibitem [{\citenamefont {Adamenko}\ \emph {et~al.}(2009)\citenamefont
  {Adamenko}, \citenamefont {Kitsenko}, \citenamefont {Nemchenko},\ and\
  \citenamefont {Wyatt}}]{Wyatt2009}%
  \BibitemOpen
  \bibfield  {author} {\bibinfo {author} {\bibfnamefont {I.~N.}\ \bibnamefont
  {Adamenko}}, \bibinfo {author} {\bibfnamefont {Y.~A.}\ \bibnamefont
  {Kitsenko}}, \bibinfo {author} {\bibfnamefont {K.~E.}\ \bibnamefont
  {Nemchenko}},\ and\ \bibinfo {author} {\bibfnamefont {A.~F.~G.}\ \bibnamefont
  {Wyatt}},\ }\bibfield  {title} {\bibinfo {title} {{Theory of scattering
  between two phonon beams in superfluid Helium}},\ }\href
  {https://journals.aps.org/prb/pdf/10.1103/PhysRevB.80.014509} {\bibfield
  {journal} {\bibinfo  {journal} {Phys. Rev. B}\ }\textbf {\bibinfo {volume}
  {80}},\ \bibinfo {pages} {014509} (\bibinfo {year} {2009})}\BibitemShut
  {NoStop}%
\bibitem [{\citenamefont {Stamper-Kurn}\ \emph {et~al.}(1998)\citenamefont
  {Stamper-Kurn}, \citenamefont {Miesner}, \citenamefont {Inouye},
  \citenamefont {Andrews},\ and\ \citenamefont {Ketterle}}]{Ketterle1998}%
  \BibitemOpen
  \bibfield  {author} {\bibinfo {author} {\bibfnamefont {D.~M.}\ \bibnamefont
  {Stamper-Kurn}}, \bibinfo {author} {\bibfnamefont {H.-J.}\ \bibnamefont
  {Miesner}}, \bibinfo {author} {\bibfnamefont {S.}~\bibnamefont {Inouye}},
  \bibinfo {author} {\bibfnamefont {M.~R.}\ \bibnamefont {Andrews}},\ and\
  \bibinfo {author} {\bibfnamefont {W.}~\bibnamefont {Ketterle}},\ }\bibfield
  {title} {\bibinfo {title} {{Collisionless and Hydrodynamic Excitations of a
  Bose-Einstein Condensate}},\ }\href
  {https://doi.org/10.1103/PhysRevLett.81.500} {\bibfield  {journal} {\bibinfo
  {journal} {Phys. Rev. Lett.}\ }\textbf {\bibinfo {volume} {81}},\ \bibinfo
  {pages} {500} (\bibinfo {year} {1998})}\BibitemShut {NoStop}%
\bibitem [{\citenamefont {Buggle}\ \emph {et~al.}(2005)\citenamefont {Buggle},
  \citenamefont {Pedri}, \citenamefont {von Klitzing},\ and\ \citenamefont
  {Walraven}}]{Walraven2005}%
  \BibitemOpen
  \bibfield  {author} {\bibinfo {author} {\bibfnamefont {C.}~\bibnamefont
  {Buggle}}, \bibinfo {author} {\bibfnamefont {P.}~\bibnamefont {Pedri}},
  \bibinfo {author} {\bibfnamefont {W.}~\bibnamefont {von Klitzing}},\ and\
  \bibinfo {author} {\bibfnamefont {J.~T.~M.}\ \bibnamefont {Walraven}},\
  }\bibfield  {title} {\bibinfo {title} {{Shape oscillations in nondegenerate
  Bose gases: Transition from the collisionless to the hydrodynamic regime}},\
  }\href {https://doi.org/10.1103/PhysRevA.72.043610} {\bibfield  {journal}
  {\bibinfo  {journal} {Phys. Rev. A}\ }\textbf {\bibinfo {volume} {72}},\
  \bibinfo {pages} {043610} (\bibinfo {year} {2005})}\BibitemShut {NoStop}%
\bibitem [{\citenamefont {Kurkjian}\ \emph
  {et~al.}(2016{\natexlab{b}})\citenamefont {Kurkjian}, \citenamefont
  {Castin},\ and\ \citenamefont {Sinatra}}]{Kurkjian2016EPL}%
  \BibitemOpen
  \bibfield  {author} {\bibinfo {author} {\bibfnamefont {H.}~\bibnamefont
  {Kurkjian}}, \bibinfo {author} {\bibfnamefont {Y.}~\bibnamefont {Castin}},\
  and\ \bibinfo {author} {\bibfnamefont {A.}~\bibnamefont {Sinatra}},\
  }\bibfield  {title} {\bibinfo {title} {{Landau-Khalatnikov phonon damping in
  strongly interacting Fermi gases}},\ }\href
  {https://doi.org/10.1209/0295-5075/116/40002} {\bibfield  {journal} {\bibinfo
   {journal} {EPL}\ }\textbf {\bibinfo {volume} {116}},\ \bibinfo {pages}
  {40002} (\bibinfo {year} {2016}{\natexlab{b}})}\BibitemShut {NoStop}%
\bibitem [{\citenamefont {Castin}(2019)}]{Castin2019}%
  \BibitemOpen
  \bibfield  {author} {\bibinfo {author} {\bibfnamefont {Y.}~\bibnamefont
  {Castin}},\ }\bibfield  {title} {\bibinfo {title} {{Bragg spectroscopy and
  pair-breaking-continuum mode in a superfluid Fermi gas}},\ }\href@noop {}
  {\bibfield  {journal} {\bibinfo  {journal} {arXiv:1911.10950v3}\ } (\bibinfo
  {year} {2019})}\BibitemShut {NoStop}%
\bibitem [{\citenamefont {Kurkjian}\ \emph {et~al.}(2020)\citenamefont
  {Kurkjian}, \citenamefont {Tempere},\ and\ \citenamefont
  {Klimin}}]{Kurkjian2020}%
  \BibitemOpen
  \bibfield  {author} {\bibinfo {author} {\bibfnamefont {H.}~\bibnamefont
  {Kurkjian}}, \bibinfo {author} {\bibfnamefont {J.}~\bibnamefont {Tempere}},\
  and\ \bibinfo {author} {\bibfnamefont {S.~N.}\ \bibnamefont {Klimin}},\
  }\bibfield  {title} {\bibinfo {title} {Linear response of a superfluid fermi
  gas inside its pair-breaking continuum},\ }\href
  {https://doi.org/10.1038/s41598-020-65371-9} {\bibfield  {journal} {\bibinfo
  {journal} {Scientific Reports}\ }\textbf {\bibinfo {volume} {10}},\ \bibinfo
  {pages} {11591} (\bibinfo {year} {2020})}\BibitemShut {NoStop}%
\bibitem [{\citenamefont {Hu}\ \emph {et~al.}(2010)\citenamefont {Hu},
  \citenamefont {Taylor}, \citenamefont {Liu}, \citenamefont {Stringari},\ and\
  \citenamefont {Griffin}}]{Griffin2010}%
  \BibitemOpen
  \bibfield  {author} {\bibinfo {author} {\bibfnamefont {H.}~\bibnamefont
  {Hu}}, \bibinfo {author} {\bibfnamefont {E.}~\bibnamefont {Taylor}}, \bibinfo
  {author} {\bibfnamefont {X.-J.}\ \bibnamefont {Liu}}, \bibinfo {author}
  {\bibfnamefont {S.}~\bibnamefont {Stringari}},\ and\ \bibinfo {author}
  {\bibfnamefont {A.}~\bibnamefont {Griffin}},\ }\bibfield  {title} {\bibinfo
  {title} {{Second sound and the density response function in uniform
  superfluid atomic gases}},\ }\href
  {https://doi.org/10.1088/1367-2630/12/4/043040} {\bibfield  {journal}
  {\bibinfo  {journal} {New Journal of Physics}\ }\textbf {\bibinfo {volume}
  {12}},\ \bibinfo {pages} {043040} (\bibinfo {year} {2010})}\BibitemShut
  {NoStop}%
\bibitem [{\citenamefont {Feynman}(1954)}]{Feynman1954}%
  \BibitemOpen
  \bibfield  {author} {\bibinfo {author} {\bibfnamefont {R.~P.}\ \bibnamefont
  {Feynman}},\ }\bibfield  {title} {\bibinfo {title} {{Atomic Theory of the
  Two-Fluid Model of Liquid Helium}},\ }\href
  {https://doi.org/10.1103/PhysRev.94.262} {\bibfield  {journal} {\bibinfo
  {journal} {Phys. Rev.}\ }\textbf {\bibinfo {volume} {94}},\ \bibinfo {pages}
  {262} (\bibinfo {year} {1954})}\BibitemShut {NoStop}%
\bibitem [{\citenamefont {Landau}(1941{\natexlab{b}})}]{Landau1941PR}%
  \BibitemOpen
  \bibfield  {author} {\bibinfo {author} {\bibfnamefont {L.}~\bibnamefont
  {Landau}},\ }\bibfield  {title} {\bibinfo {title} {{Theory of the
  Superfluidity of Helium II}},\ }\href
  {https://doi.org/10.1103/PhysRev.60.356} {\bibfield  {journal} {\bibinfo
  {journal} {Phys. Rev.}\ }\textbf {\bibinfo {volume} {60}},\ \bibinfo {pages}
  {356} (\bibinfo {year} {1941}{\natexlab{b}})}\BibitemShut {NoStop}%
\end{thebibliography}
\end{document}


\title{
Supplemental Material for \\
``Effects of quantum fluctuations 
on the low-energy collective modes \\
of two-dimensional superfluid Fermi gases from 
the BCS to the Bose Limit''
}
%
%

\author{Senne Van Loon}
\email{Senne.Van\_Loon@colostate.edu}
\affiliation{School of Physics, Georgia Institute of Technology, Atlanta, Georgia 30332, USA}
\affiliation{TQC, Universiteit Antwerpen, Universiteitsplein 1, B-2610
                Antwerpen, Belgi\"e}
\author{C. A. R. S\'a de Melo}
\affiliation{School of Physics, Georgia Institute of Technology, Atlanta, Georgia 30332, USA}
\date{\today}

\maketitle

In this supplemental material, we provide details of a few fundamental points
that were made in the main text. We correct some myths that have been
perpetuated in the literature. The first one is the issue of damping of the
low-energy collective mode at zero temperature $(T = 0)$ in 2D. The second point
is that the poles of the structure factor at $T = 0$ correspond to the broken
U(1) collective mode and not to Landau's first sound. The third issue is that
the sound velocity of the broken U(1) collective mode at $T = 0$ is always
quantitatively smaller than the isothermal or isentropic sound velocities, which
are often used as approximations for Landau's first sound.

\section{Effects of Damping: 2D versus 3D}
\label{sec:effects-of-damping}

In our discussion about damping, we limit ourselves to the case of neutral
$s$-wave superfluids with short-ranged interactions at $T = 0$, since we only
computed the broken U(1) collective mode dispersion for this case.

There is a long-standing result in the literature stating that the broken U(1)
collective mode at $T = 0$ always exhibits damping at sufficiently weak
interactions due to pair-breaking~\cite{Popov1991}. This expectation is based on
linear extrapolations of the broken U(1) collective modes $\omega_{\rm col}
({\bf q}) = c \vert {\bf q} \vert$ into higher energies, suggesting that the
collective mode energy crosses the two-quasiparticle continuum threshold $2
\vert \Delta \vert$ in 2D and 3D~\cite{Popov1991, Combescot2006,Kurkjian2016}
for sufficiently weak interactions, where Cooper pairs may be easily broken.
Such approximation suggests that for $\vert {\bf q} \vert > 2 \vert \Delta
\vert/c$, where $c$ is the speed of sound, damping occurs. This result is
strongly based on two assumptions. First, that one has a BCS superfluid, and
second that the linear approximation $\omega_{\rm col} ({\bf q}) = c \vert {\bf
q} \vert $ for the collective mode dispersion still holds for $\vert {\bf q}
\vert > 2 \vert \Delta \vert/c$.

Recent investigations~\cite{Kurkjian2016} in 3D included higher order
corrections in $\vert {\bf q} \vert$ leading to 
$
\omega_{\rm col} ({\bf q}) 
= 
c \vert {\bf q} \vert 
+ \widetilde{\gamma} \vert {\bf q} \vert^3
+ 
\widetilde{\eta} \vert {\bf q} \vert^5
$
and showed that indeed there is a pair breaking regime as naively expected from
the linear extrapolation. However, the the value of the wavenumber $\vert {\bf q} \vert$
where the pair-breaking threshold is reached, gets pushed to higher values.
This reduction of the damping region due to pair-breaking was attributed to the
change in concavity of the dispersion relation in the BCS
regime~\cite{Kurkjian2016}.

In 2D, our results show that the inclusion of higher order corrections has an
even stronger effect, that is, the change in concavity completely suppresses the
decay of the collective mode into the two-quasiparticle continuum for all ${\bf
q}$ and all interaction strengths. Therefore, the naive
expectation~\cite{Popov1991}, based on the linear extrapolation, that the broken
U(1) collective mode always decays into two quasiparticles at sufficiently large
wavevectors and energies for weak interactions is wrong. The key insight on
damping provided by our work is that the strong change in the concavity of
$\omega_{\rm col} ({\bf q})$, that avoids the two-quasiparticle threshold
altogether, is due to the existence of stable two-body bound states for
arbitrarily weak $s$-wave interactions in 2D. This means that the decay mechanism
of the broken U(1) collective modes into two quasiparticles is completely
suppressed in 2D by the ubiquitous presence of two-body bound states, in sharp
contrast with the 3D situation where decay into the two-quasiparticle continuum
still occurs. 

Having shown that decay of the broken U(1) collective mode into
two-quasiparticles (pair breaking) is forbidden in 2D at $T = 0$, a natural
question to ask is: Are there additional decay (damping) mechanisms that may
arise? The answer to this question is simple. To Gaussian order in the order
parameter fluctuations, the only other decay mechanism is Landau damping, which
occurs at $T \ne 0$ and is due to quasiparticle-quasihole processes, both in
2D and 3D. Since we are at $T = 0$, the Landau damping mechanism is thermally
suppressed and does not emerge in our calculation. 

Beyond Gaussian order there are possible damping mechanisms such as those
proposed by Beliaev~\cite{Beliaev1958}, where a collective mode (phonon) can
decay into two collective modes (phonons) with lower energies, or by
Landau-Khalatnikov~\cite{Landau-Khalatknikov1949}, where two collective modes
(phonons) can decay into two other collective modes (phonons). In the absence of
disorder, the damping of low-energy collective modes (phonons) is determined by
collective-mode-collective-mode (phonon-phonon) interactions that conserve
momentum and energy. Therefore, the damping rate depends strongly on the
curvature of the collective mode (phonon) dispersion relation~\cite{Wyatt1992,
Wyatt2009}. The concavity change in the collective mode dispersion, from concave
(BCS) to convex (Bose), can lead to different outcomes. Ultracold atom Fermi
superfluids are believed to be in the collisionless regime where $\omega_{\rm
col} ({\bf q}) \tau \gg 1$, with $\tau$ being a characteristic collision
time~\cite{Ketterle1998,Walraven2005}. Within the concave region of $\omega_{\rm
col} ({\bf q})$, Beliaev processes are kinematically suppressed, but
Landau-Khalatnikov damping can exist~\cite{Kurkjian2016EPL}. In the convex
regime, the two processes may contribute to damping of the collective modes.
However, we do not address effects beyond Gaussian order in our work.

Having clarified the issue of damping of collective modes in 2D at $T = 0$, we
shown next that the poles of the structure factor are not given by Landau's
first sound at $T = 0$, but rather by the broken U(1) collective mode that we
calculate.

\section{Density-density response}
\label{sec:density-density-response}

In this section, we show that the poles of the density-density correlation
function are identical to the poles of the modulus-phase fluctuation matrix at
zero temperature. Therefore, the collective modes detected by the dynamical
structure factor in the superfluid state at $T = 0$ are the ones that arise from
the spontaneously broken U(1) symmetry. 

We obtain the density-density correlation function by adding a source term 
\begin{equation}
S_{J} ({\bar \psi}_s, \psi_s)  = 
\int_0^{\beta} d\tau \int d^{D}{\bf x}
J(x) \sum_{s}{\bar \psi}_s (x) \psi_s (x)
\end{equation}
to the action ${\cal S}$ corresponding to the Hamiltonian density ${\cal H}$
given in Eq.~(1) of the main text. The source field $J(x)$, with dimensions of
energy, couples to the local density  $\rho (x) = \sum_{s} {\bar \psi}_s (x)
\psi_s (x)$, where $x = ({\bf x}, \tau)$. Integrating the fermions leads to the
second-order contribution
\begin{equation}
S_2 = 
\frac{\beta}{2}
\sum_{q} {\boldsymbol \Phi}^{\rm T} (-q)
{\bf M} (q) {\boldsymbol \Phi} (q),
\end{equation}
where the vector ${\boldsymbol \Phi}^{\rm T} = (\eta_{\rm R}, \eta_{\rm I}, J)$
is three-dimensional, ${\rm T}$ means transposition, $\eta_{\rm R}$ and
$\eta_{I}$ are the real and imaginary parts of the fluctuations in the order
parameter, and $q = ({\bf q}, i\omega)$. The matrix correlating ${\boldsymbol
\Phi}^{\rm T}$ and $\boldsymbol{\Phi}$ is
\begin{equation}
{\bf M} 
= 
\begin{pmatrix}
M_{\rm RR} & M_{\rm RI} & M_{\rm RJ} \\
M_{\rm IR} & M_{\rm II} & M_{\rm IJ} \\
M_{\rm JR} & M_{\rm JI} & M_{\rm JJ}
\end{pmatrix},
\end{equation}
has dimensions of inverse energy, and the first $2\times 2$ block represents the
order parameter fluctuation matrix 
\begin{equation}
{\widetilde {\bf M}} 
= 
\begin{pmatrix}
M_{\rm RR} & M_{\rm RI} \\
M_{\rm IR} & M_{\rm II} \\
\end{pmatrix}.
\end{equation}
Integration over the order parameter fluctuations produces the $JJ$ action
\begin{equation}
S_{2 {\rm JJ}}
= 
\frac{\beta}{2}
\sum_q 
J(-q) 
\chi_{\rho \rho} (q)
J(q), 
\end{equation}
that is second order in the source $J$, leading to the density-density
correlation function 
\begin{equation}
\label{eqn:chi-rho-rho}
\chi_{\rho\rho} (q)   
=
M_{\rm JJ} (q) 
- 
\frac{D_{\rm JJ} (q)}{\det {\widetilde {\bf M}} (q)},
\end{equation}
that has dimensions of inverse energy.
The term $M_{\rm JJ} (q)$ has no poles for low frequencies at $T = 0$, thus the
poles of $\chi_{\rho\rho} (q)$ are determined solely by the integer-order zeros
of $\det {\widetilde {\bf M}} (q) = 0$, with $D_{\rm JJ} (q)$ playing a  role in
determining the residue of the poles obtained via the analytic continuation $q =
({\bf q}, i\omega) \to ({\bf q}, \omega + i\delta)$.

This expression shows that the low-frequency poles of $\chi_{\rho \rho} ({\bf
q}, \omega + i\delta)$ are the same as those from the inverse fluctuation matrix
${\widetilde {\bf M}}^{-1}$, which describes the order parameter fluctuations.
The result in Eq.~(\ref{eqn:chi-rho-rho}) is mathematically the same as that
found in the literature~\cite{Combescot2006, Castin2019, Kurkjian2020} using
standard linear response theory. The major advantage of our approach is that it
shows explicitly that the low-frequency poles of $\chi_{\rho\rho} ({\bf q},
\omega + i\delta)$ and of the order parameter fluctuation matrix ${\widetilde
{\bf M}} ({\bf q}, \omega + i\delta)$ are exactly the same. The explicit
expressions of $M_{\rm JJ} (q)$ and $D_{\rm JJ} (q)$ are given in the Appendix.

At $T = 0$, the density-density structure factor is
\begin{equation}
\label{eqn:dynamical-structure-factor}
S({\bf q}, \omega) = 
-\frac{1}{\pi}
{\rm Im} \chi_{\rho \rho} ({\bf q}, \omega + i\delta),
\end{equation}
implying that the peaks of $S({\bf q}, \omega)$ correspond to the poles
$\omega_{\rm col} ({\bf q})$ of $\chi_{\rho\rho} ({\bf q}, \omega + i\delta)$,
and are therefore a manifestation of the broken U(1) symmetry of the superfluid
state. This proves that what $S({\bf q}, \omega)$ measures at $T = 0$ is
microscopically equivalent to the collective modes revealed in the order
parameter fluctuation matrix discussed in the main text. 

There have been claims in the literature that Landau's two-fluid model can be
used to describe collective modes of Fermi superfluids detected in $S({\bf q},
\omega)$~\cite{Griffin2010}, provided that the {\it normal} and {\it superfluid}
components are in local hydrodynamic equilibrium. Such phenomenological
construction requires short collision times between excitations in the {\it
normal} fluid and an artificial separation between {\it normal} and {\it
superfluid} components.

The artificiality of Landau's phenomenological theory was thoroughly discussed
by Feynman~\cite{Feynman1954}, who criticized the approach, because physically
there is only one fluid and not two. What happens microscopically is that the
fluid changes its properties below the critical temperature. Nevertheless,
practitioners of Landau's approach like to justify the separability of a single
fluid into a {\it normal} and {\it superfluid} component by assuming that the
frequency of collective modes $\omega_{\rm col} ({\bf q})$ must be much smaller
than a typical relaxation rate $1/\tau$, that is, $\omega_{\rm col} ({\bf q})
\tau \ll 1$. Furthermore, it has been argued that local hydrodynamic equilibrium
in ultracold atoms is very difficult to achieve because the scattering length
and densities are not sufficiently  
large~\cite{Griffin2010}. In the case of ultracold fermions, it was assumed that
this may be possible near unitarity, but apparently nowhere
else~\cite{Griffin2010}. However, experimentally, these systems seem to be in
the colisionless regime $\omega_{\rm col} ({\bf q}) \tau \gg 1$ as $T \to
0$~\cite{Ketterle1998, Walraven2005, Kurkjian2016EPL}.

The question then arises: Is the broken U(1) collective mode at $T = 0$,
revealed by the dynamical structure factor, in any way related to Landau's first
sound? This issue is addressed next.

\section{Broken U(1) vs. Landau First Sound} 
\label{sec:broken-U1-landau-first-sound}

We established unequivocally that the low-frequency collective modes extracted
experimentally from $S({\bf q}, \omega)$ at $T = 0$ are indeed the broken U(1)
symmetry modes discussed in the main text. At unitarity, Landau's first and
second sounds are assumed to be collective modes that arise from the poles of
$S({\bf q}, \omega)$~\cite{Griffin2010} provided that $\omega_{\rm col} ({\bf
q})\tau \ll 1$, but often, in conversations, Landau's first sound is identified
as the collective mode that is measured from $S({\bf q}, \omega)$ throughout the
BCS-Bose evolution without regard to the required conditions for reaching the
hydrodynamic limit. For us, this is a serious concern. 

Keeping the previous statements in mind, it is of fundamental importance to ask
the question: Are the broken U(1) and Landau's first sound the same at $T = 0$?
Our answer is that these sound velocities are not the same either conceptually
or quantitatively. Landau's first sound does not invoke broken U(1) symmetry in
a fluid that changes its properties below the critical temperature $T_c$, rather
it relies on the phenomenological construction of two fluids with different
densities, where the superfluid density vanishes at and above $T_c$. In Landau's
approach there is no broken symmetry explicitly invoked.

According to Landau's original work~\cite{Landau1941USSR}, the first sound
velocity $u_1$ is
\begin{equation}
\label{eqn:first-sound}
u_1^2 =
\frac{b}{2} +
\sqrt{ \left(\frac{b}{2}\right)^2 - g},
\end{equation}
while the second sound velocity $u_2$ is 
\begin{equation}
\label{eqn:first-sound}
u_2^2 =
\frac{b}{2} -
\sqrt{ \left(\frac{b}{2}\right)^2 - g}.
\end{equation}
The first coefficient appearing above is
\begin{equation}
b = 
u_S^2
=\left(\frac{\partial P}{\partial \rho}\right)_S, 
\end{equation}
representing the square of the isentropic sound velocity $u_S$, where $P$ is the
pressure, $\rho$ is the mass density and $S$ is the entropy. The second
coefficient is 
\begin{equation}
g = \frac{TS^2}{C_V} \frac{\rho_s}{\rho_n} 
\left(\frac{\partial P}{\partial\rho}\right)_T,
\end{equation}
having dimensions of velocity to the fourth power, where $T$ is temperature,
$C_V$ is the heat capacity at constant volume $V$, $\rho_s$ is the superfluid
density, and $\rho_n$ is the normal density. The first few factors of $g$ can be
grouped into a squared velocity
\begin{equation}
u_e^2 = 
\frac{TS^2}{C_V} \frac{\rho_s}{\rho_n}, 
\end{equation}
while the last factor in $g$ is 
\begin{equation}
u_T^2 = 
\left(\frac{\partial P}{\partial\rho}\right)_T,
\end{equation}
where $u_T$ is the isothermal sound velocity.

In the regions of phase space where the superfluid has a small coefficient
of thermal expansion $\alpha$, the first sound velocity is approximately the
isentropic sound velocity, that is, $u_1 \approx u_S$~\cite{Landau1941PR,
Landau1941USSR} since $g \ll (b/2)^2$ in Eq.~(\ref{eqn:first-sound}). The
thermodynamic relation $u_S^2 = \gamma u_T^2$, where $\gamma = C_P/C_V$ is the
ratio between the heat capacities at constant pressure $(C_P)$ and at constant
volume $(C_V)$, guarantees that $u_S \ge u_T$ since $\gamma \ge 1$.  

The microscopic dynamical structure factor $S({\bf q}, \omega)$ can only be
related to the phenomenological first and second sounds in the hydrodynamic
regime~\cite{Martin1964, Martin1965} where $\omega \tau \ll 1$. Here, $\tau$ is
a temperature dependent relaxation time, which is usually a high power of
inverse temperature~\cite{Martin1964,Martin1965}, when the artificial separation
between normal and superfluid components may have a microscopic justification, at
least for bosonic systems~\cite{Martin1964, Martin1965}. The inverse relaxation
time can be parametrized as $\tau^{-1} = \omega_0 (T/T_c)^\ell$, where
$\omega_0$ is a characteristic relaxation frequency of the system, $T_c$ is the
superfluid transition temperature and $\ell$ is typically in the interval $4 \le
\ell \le 6$ in 3D or $3 \le \ell \le 5$ in 2D. The hydrodynamic condition
requires that $\omega \ll \omega_0 (T/T_c)^\ell$. Strictly at $T = 0$, the
hydrodynamic limit defined above does not exist. Therefore, at $T = 0$, our
microscopic collective mode describing the broken U(1) symmetry cannot be
Landau's first sound. Using the Hohenberg-Martin terminology, our microscopic
broken U(1) collective mode arises as a pole of $S({\bf q},\omega)$ in the
collisionless regime $\omega \tau \gg 1$, which seems to be the experimental
situation in ultracold fermions~\cite{Ketterle1998,Walraven2005,
Kurkjian2016EPL}.

The question then becomes: Is there a relationship between Landau's first sound
and the broken U(1) collective mode at $T = 0$? In the next section we will
address this question via the compressibility sum rule. 

\section{Compressibility Sum Rule}
\label{sec:compressibility-sum-rule}

The dynamical structure factor in Eq.~(\ref{eqn:dynamical-structure-factor})
satisfies the compressibility sum rule
\begin{equation}
\label{eqn:compressibility-sum-rule}
\lim_{{\bf q} \to 0} 
\int_{-\infty}^{+\infty} d\omega
\frac{S({\bf q}, \omega)}{\omega} = 
\frac{1}{2m u_T^2}.
\end{equation}
Using the explicit expression for $\chi_{\rho \rho} (q)$ in
Eq.~(\ref{eqn:chi-rho-rho}) we can show that 
\begin{equation}
\label{eqn:compressibility-U-1}
\lim_{{\bf q} \to 0} 
\int_{-\infty}^{+\infty} d\omega
\frac{S({\bf q}, \omega)}{\omega} = 
\frac{1}{2m c^2}
-
\frac{
\vert \Delta \vert}
{\mu^2 + \vert \Delta \vert^2}
F(\vert\Delta\vert/\mu),
\end{equation}
where $F(\vert \Delta \vert/\mu)$ is a strictly positive dimensionless function
that depends only on the ratio $\vert \Delta \vert/\mu$. The first contribution
in Eq.~(\ref{eqn:compressibility-U-1}) comes from the pole of $S ({\bf q},
\omega)$, while the one containing $ F(\vert\Delta\vert/\mu)$ arises from the
background. Combining Eqs.~(\ref{eqn:compressibility-sum-rule})
and~(\ref{eqn:compressibility-U-1}) we obtain the relation
\begin{equation}
\left( \frac{u_T}{c}\right)^2
= 
1 +
2m u_T^2 \frac{
\vert \Delta \vert}
{\mu^2 + \vert \Delta \vert^2}
F(\vert\Delta\vert/\mu),
\end{equation}    
indicating that the isothermal sound velocity $u_T$ is always larger than the
broken U(1) sound velocity $c$. In general, $u_T \ge c$, where the equality sign
only occurs for $\vert \Delta\vert/\mu \to 0$, when the effects of the broken
U(1) symmetry are negligible (extreme BCS regime or the deep Bose limit) or for
$\vert \Delta \vert = 0$ strictly, when there is no broken U(1) symmetry. 

Since the isentropic sound velocity is $u_S = \gamma u_T$, then the inequality
$u_S \ge u_T \ge c$ holds. If we assume that the hydrodynamic limit can be
reached in the limit of $T \to 0$, and take $u_1 \approx u_S$ as argued for
$^4{\rm He}$~\cite{Landau1941PR} and for unitary Fermi
superfluids~\cite{Griffin2010}, then it is clear that the first sound $u_1$ is
not quantitatively the same as the broken U(1) sound $c$, but in the limit of $T
\to 0$ there is a potential relationship between the two conceptually distinct
quantities. However, we would like to emphasize that at $T = 0$ the poles
revealed by $S({\bf q}, \omega)$ arise from the broken U(1) collective modes and
not from Landau's first sound. 

\begin{figure}
\centering 
\includegraphics{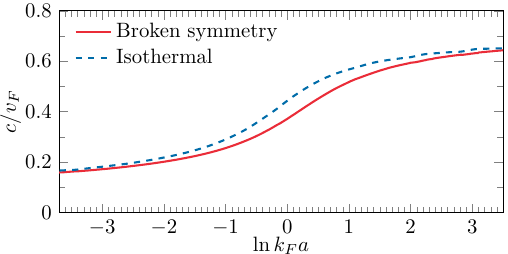}
\caption{
The sound velocity of the broken U(1) symmetry ($c$, solid red) and the
isothermal sound velocity ($u_T$, dashed blue) versus the coupling parameter
$\ln k_F a$.
}
\label{fig:one}
\end{figure}

In Fig.~\ref{fig:one}, we plot $c$ and $u_T$ to illustrate the quantitative
difference between the two sound velocities. Notice that $u_T$ is always larger
than $c$ throughout the BCS-Bose evolution, and thus, $u_1 \approx u_S$ is also
greater than $c$. The two quantities, $u_T$ and $c$, become only asymptotically
the same when $\vert \Delta \vert/\mu \to 0$, that is, either in the extreme BCS
regime or in the deep Bose limit. At $T = 0$ and with $\gamma > 1$, this implies
generally that $u_1 \approx u_S > c$.

The discussion above also illuminates the difference between the experimental
measurement of the isentropic sound velocity $u_S$~\cite{Henning2020}, where
phase imprinting produces a pressure wave that propagates through the sample at
constant entropy, and the experimental measurement of the broken-U(1) sound
velocity $c$, which we calculate at $ T = 0$, revealed by the dynamical
structure factor~\cite{Lompe2021}. Measurements of $S({\bf q}, \omega)$ have not
been performed at momentum modulus lower than $\vert {\bf q} \vert/k_F \le
0.14$, but the broken-U(1) sound velocity~\cite{Lompe2021} appears to compare
well with its isentropic counterpart~\cite{Henning2020}. We would like to
emphasize that a more accurate measurement of $c$ should reveal that it is not
only conceptually distinct from $u_S$, but also quantitatively different,
obeying the relation $c \le u_S$ discussed above.

\section{Appendix}
\label{sec:appendix}

In this appendix, we give explicit expressions for the terms appearing in
$\chi_{\rho\rho} (q)$ given in Eq.~(\ref{eqn:chi-rho-rho}).

The first term is 
\begin{equation}
M_{\rm J J} (q, i\omega)
= 
\sum_{\bf k} 
\frac{U_{\rm JJ}
({\bf k}_{+}, {\bf k}_{-})}{(i\omega)^2 - 
(E_{{\bf k}_{+}} + E_{{\bf k}_{-}})^2},
\end{equation}
where the coherence factor 
is 
\begin{equation}
U_{\rm JJ}
({\bf k}_{+}, {\bf k}_{-})
= 
\left[
 E_{{\bf k}_{+}} E_{{\bf k}_{-}}
 -
 \xi_{{\bf k}_{+}} \xi_{{\bf k}_{-}}
 + \vert \Delta_0 \vert^2 
 \right]
 L ({\bf k}_{+}, {\bf k}_{-}),
\end{equation}
with coefficient
\begin{equation}
L ({\bf k}_{+}, {\bf k}_{-})
=
\frac{ E_{{\bf k}_{+}} + E_{{\bf k}_{-}}}
{E_{{\bf k}_{+}} E_{{\bf k}_{-}}
}.
\end{equation}
The second term, introduced in Eq.~(\ref{eqn:chi-rho-rho}), is
\begin{equation}
D_{\rm JJ} ({\bf q}, i\omega)   
= 
M_{\rm RJ}^2 M_{\rm II}
+
M_{\rm IJ}^2 M_{\rm RR}
- 
2 M_{\rm RJ} M_{\rm RI} M_{\rm IJ}.
\end{equation}
Each term appearing in $D_{\rm JJ}$ is listed below. The RJ matrix element is 
\begin{equation}
M_{\rm RJ} ({\bf q}, i\omega) = 
\sum_{\bf k} 
\left[
\frac{U_{\rm RJ}
({\bf k}_{+}, {\bf k}_{-})}{(i\omega)^2 - 
(E_{{\bf k}_{+}} + E_{{\bf k}_{-}})^2}
\right],
\end{equation}
with the coherence factor
\begin{equation}
U_{\rm RJ}
({\bf k}_{+}, {\bf k}_{-})
= 
i\omega \vert \Delta_0 \vert
 L ({\bf k}_{+}, {\bf k}_{-}).
\end{equation}
The II matrix element is 
\begin{equation}
 M_{\rm II} ({\bf q}, i\omega) 
 = 
 \sum_{\bf k} 
 \left[
 \frac{1}{E_{\bf k}}
 + \frac{
 U_{\rm II} ({\bf k}_{+}, {\bf k}_{-})}
{(i\omega)^2 - 
(E_{{\bf k}_{+}} + E_{{\bf k}_{-}})^2}
\right],
\end{equation}
with coherence factor
\begin{equation}
 U_{\rm II} ({\bf k}_{+}, {\bf k}_{-})
 = 
 \left[
 E_{{\bf k}_{+}} E_{{\bf k}_{-}}
 +
 \xi_{{\bf k}_{+}} \xi_{{\bf k}_{-}}
 - \vert \Delta_0 \vert^2 
 \right]
 L ({\bf k}_{+}, {\bf k}_{-}).
\end{equation}
The IJ matrix element is 
\begin{equation}
 M_{\rm IJ} ({\bf q}, i\omega) 
 = 
 \sum_{\bf k} 
 \frac{
 U_{\rm IJ} ({\bf k}_{+}, {\bf k}_{-})}
{(i\omega)^2 - 
(E_{{\bf k}_{+}} + E_{{\bf k}_{-}})^2},
\end{equation}
where the coherence factor is
\begin{equation}
 U_{\rm IJ} ({\bf k}_{+}, {\bf k}_{-})
 = 
 \vert \Delta_0 \vert (\xi_{{\bf k}_+} + \xi_{{\bf k}_-})
 L ({\bf k}_{+}, {\bf k}_{-}).
\end{equation}
The RR matrix element is 
\begin{equation}
 M_{\rm RR} ({\bf q}, i\omega) 
 = 
 \sum_{\bf k} 
 \left[
 \frac{1}{E_{\bf k}}
 + \frac{
 U_{\rm RR} ({\bf k}_{+}, {\bf k}_{-})}
{(i\omega)^2 - 
(E_{{\bf k}_{+}} + E_{{\bf k}_{-}})^2}
\right],
\end{equation}
where the coherence factor is
\begin{equation}
 U_{\rm RR} ({\bf k}_{+}, {\bf k}_{-})
 = 
 \left[
 E_{{\bf k}_{+}} E_{{\bf k}_{-}}
 +
 \xi_{{\bf k}_{+}} \xi_{{\bf k}_{-}}
+ \vert \Delta_0 \vert^2 
 \right]
 L ({\bf k}_{+}, {\bf k}_{-}).
\end{equation}
The last matrix element needed is
\begin{equation}
 M_{\rm RI} ({\bf q}, i\omega) 
 = 
 \sum_{\bf k} 
 \frac{
 U_{\rm RI} ({\bf k}_{+}, {\bf k}_{-})}
{(i\omega)^2 - 
(E_{{\bf k}_{+}} + E_{{\bf k}_{-}})^2},
\end{equation}
where the coherence factor is
\begin{equation}
 U_{\rm RI} ({\bf k}_{+}, {\bf k}_{-})
 = 
\frac{i\omega ( \xi_{{\bf k}_+} 
 E_{{\bf k}_{-}}
 +
 \xi_{{\bf k}_-} 
 E_{{\bf k}_{+}})}{E_{{\bf k}_{+}} E_{{\bf k}_{-}}}.
\end{equation}
The matrix elements given in this appendix are necessary to obtain the results
discussed above.



%